\documentclass[twocolumn]{autart}    
\usepackage[english]{babel}                            
\usepackage{ amsmath, amssymb, amsfonts}         
\usepackage{graphicx}                                 
\usepackage{color}    
\usepackage{mathrsfs}     
\usepackage{cancel}                            
\usepackage{enumerate}
\usepackage{enumerate}

\def\Real{\mathbb{R}}

\newcommand{\TwoOne}[2]
{\begin{bmatrix}
{#1} \\
{#2}
\end{bmatrix}
}
\newcommand{\TwoTwo}[4]
{\begin{bmatrix}
{#1} & {#2} \\
{#3} & {#4}
\end{bmatrix}
}
\newcommand{\RE}{\ensuremath{\mathrm{Re}}}
\newcommand{\Lone}{\mathcal{L}_{\rm 1}}
\newcommand{\LTE}{\mathcal{L}_{\rm 2e}}
\newcommand{\LT}{\mathcal{L}_{\rm 2}}
\newcommand{\Ltwo}{\mathcal{L}_{\rm 2}}

\begin{document}

\begin{frontmatter}

\title{On the Necessity and Sufficiency of the Zames-Falb Multipliers for Bounded Operators \thanksref{footnoteinfo}} 

\thanks[footnoteinfo]{This paper was not presented at any IFAC 
meeting. Corresponding author  Lanlan Su.}

\author[1]{Sei Zhen Khong}\ead{szkhongwork@gmail.com},    
\author[2]{Lanlan Su}\ead{ls499@leicester.ac.uk}  

\address[1]{Independent researcher}  

\address[2]{School of Engineering, University of Leicester, Leicester, LE1 7RH, UK}        


\begin{abstract}                          
  This paper analyzes the robust feedback stability of a single-input-single-output stable linear time-invariant (LTI) system against four different
  classes of nonlinear systems using the Zames-Falb multipliers.  The contribution is fourfold.  Firstly, we present a  generalised S-procedure lossless theorem that involves a countably infinite number of quadratic forms. Secondly, we identify a class of uncertain systems
  over which the robust feedback stability implies the existence of an appropriate Zames-Falb multiplier based on the generalised S-procedure lossless theorem. Meanwhile, we show that the existence of such a Zames-Falb multiplier is sufficient for the robust feedback stability over a smaller class of uncertain systems. Thirdly, when restricted to be
  static (a.k.a. memoryless), the second class of systems coincides with the class of sloped-restricted monotone nonlinearities, and the classical result
  of using the Zames-Falb multipliers to ensure feedback stability is recovered.  Lastly, when restricted to be LTI, the second class is demonstrated
  to be a subset of the third, and the existence of a Zames-Falb multiplier is shown to be sufficient but not necessary for the robust feedback
  stability.
\end{abstract}

\end{frontmatter}

\section{Introduction}

\begin{figure}
\begin{center}
\includegraphics[height=4cm]{{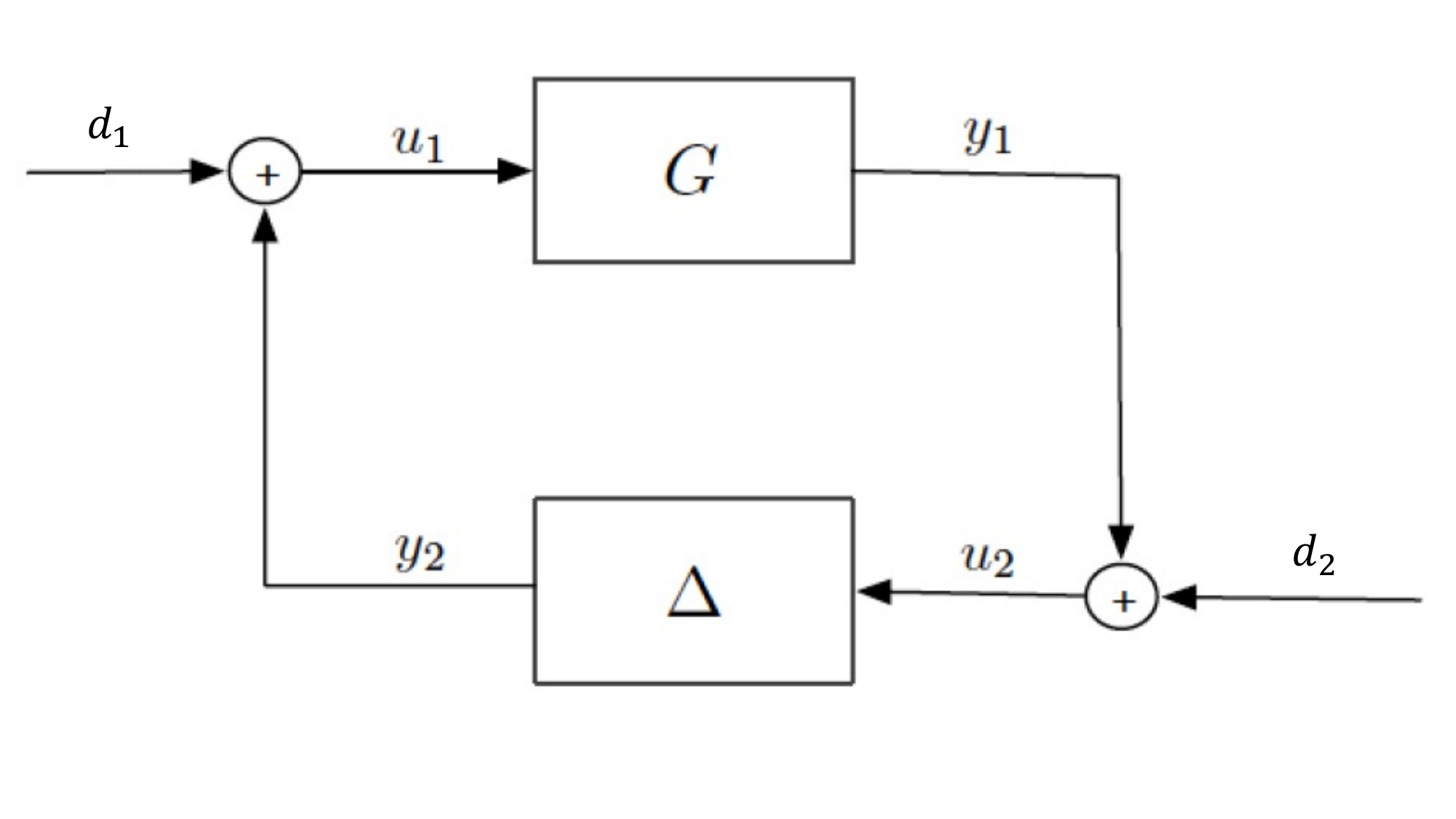}}    

 \end{center} 
 \caption{Standard feedback configuration }     
 \label{fig:1}                            
\end{figure}

The robust stability analysis of the feedback interconnection of a single-input-single-output stable linear time-invariant (LTI) system $G$ and a
nonlinear system $\Delta$ belonging to a specified uncertainty class, as depicted in Figure \ref{fig:1}, is a fundamental object of study in the field
of control theory. It is often termed absolute stability analysis in the nonlinear systems literature \cite{Pop61,Yak82}. When the class of $\Delta$'s
considered is static (i.e. memoryless) and sector bounded, a variety of multiplier-based methodologies including the renowned circle criterion and
Popov criterion \cite{khalil2002nonlinear} have been proposed to establish the input-output feedback stability. For static, time-invariant, and
monotone $\Delta$'s, the Zames-Falb multipliers are currently the most general class of time-invariant multipliers known for the study of these
feedback systems \cite{carrasco2013equivalence}.

The Zames-Falb multipliers were first introduced by O'Shea in \cite{o1966combined,o1967improved} (see \cite{carrasco2016zames} for a survey), and
formalized later by Zames and Falb in \cite{zames1967stability}. Unfortunately, its applicability was limited due to computational
constraints. Motivated by the rapid development of computational capability in the 80s, the multiplier-based theory has regained the interests of
researchers. Importantly, the seminal work \cite{megretski1997system} demonstrates that the Zames-Falb multipliers fit nicely to the framework of
integral quadratic constrains (IQCs); see also~\cite{Kho22}. A recent research line on numerically searching the Zames-Falb multipliers satisfying a
mixture of time and frequency domain conditions can be located in (\cite{carrasco2014lmi,turner2019analysis,turner2009existence}).  The phase
limitation of the Zames-Falb multipliers has been studied in (\cite{megretski1995combining,jonsson1996stability,wang2017phase}), thereby providing a
interesting perspective from which to understand the Zames-Falb multipliers.

It is well known~\cite{zames1967stability} that the robust feedback stability of an LTI system $G$ to static monotone nonlinearities $\Delta$'s with a
slope restriction $b$ can be ensured by the existence of an appropriate Zames-Falb multiplier $M(j\omega)$ satisfying
\begin{equation}\label{eq:intro}
  \RE \left\lbrace  M(j\omega) (b^{-1}-G(j\omega))\right\rbrace>0,\forall \omega\in \mathbb{R}\cup \infty.
\end{equation}
In \cite{carrasco2016zames,wang2017phase}, it is conjectured by Carrasco that if there is no appropriate Zames-Falb multiplier, then the feedback
system is not robustly stable. The Carrasco's conjecture remains unsolved, i.e. it is unclear whether condition \eqref{eq:intro} is necessary for the
robust feedback stability over the class of static monotone $\Delta$'s. Partial discrete-time results on the conjecture can be found
  in~\cite{SeiCar21,ZCH20}. The purpose of this work is to investigate both the necessity and sufficiency of the Zames-Falb multipliers in the
continuous time over different uncertainty classes of $\Delta$'s, with the goal of enhancing our understanding of the conservatism of the Zames-Falb
multipliers in the robust stability analysis of the feedback system in Figure \ref{fig:1}. The necessity of various robust stability conditions has
been studied in control over the years, since the pioneering work on that of the small-gain theorem for LTI systems~\cite{Doy84,ZDG96}. Related
converse results for IQCs can be found in~\cite{KhoSch18,KhoKao21}.

This paper is concerned with the robust stability analysis of the feedback interconnection shown in Figure \ref{fig:1} over four uncertainty classes
of $\Delta$'s using the Zames-Falb multipliers. While one might be inclined to yearn for a condition that guarantees the robust feedback stability
over as large a class of $\Delta$'s as possible, this is unwise from the perspective of establishing the necessity of the condition. In fact, there is
a subtle trade-off between the size of the class of $ \Delta$'s and the strictness of the condition on $G$ when it comes to robust stability, the
latter of which obviously affects its necessity. Understandably, a larger class of $ \Delta$'s leads to a stricter condition on $G$ in order to ensure
robust stability.  This is well illustrated by our results involving classes of $\Delta$'s of different sizes.  Firstly, we identify a set of
nonlinear dynamic $\Delta$'s and show that the existence of a Zames-Falb multiplier satisfying condition \eqref{eq:intro} is implied by the robust
stability against this uncertainty set. This is established using the  generalised S-procedure lossless theorem that is presented in Section \ref{sec: s-lemma}. Moreover, we show that the existence of such a Zames-Falb multiplier is sufficient for the robust feedback stability over a smaller class of uncertain systems. Secondly, when this set is restricted to be static, it coincides with the class of static monotone
nonlinearities, and the existence of a Zames-Falb multiplier is sufficient for the robust stability whereas its necessity remains unknown. Thirdly,
when the same set is restricted to consist of only LTI dynamics, it is shown to be a subset of the class of static monotone nonlinearities, and the
existence of a Zames-Falb multiplier is sufficient but not necessary to establish the uniform stability. For the ease of exposition, the main results
centered around monotone nonlinearities $\Delta$'s is presented in Section \ref{sec:3}, and Section \ref{sec:4} is dedicated to its extension to the
general two-sided slope restrictions. Specifically, in Section~\ref{sec:4}, the sufficiency direction considered
  in~\cite{zames1967stability} is importantly generalised in order to make the condition consistent with the necessity direction.  A crucial part of our results shows that the `monotonicity' in the set of monotone static time-invariant
nonlinearities is completely characterizable via the Zames-Falb IQCs. Therefore, while there exists a larger class of IQCs that such nonlinearities
satisfy~\cite{KulSaf02}, using them does not deliver additional benefits as far as establishing robust feedback stability is concerned.

The remainder of this section sets up the notation and mathematical preliminaries to the rest of the paper. Section \ref{sec: s-lemma} generalises  the  S-procedure lossless theorem  from~\cite{MegTre93} to involve a countably infinite number of quadratic forms. This generalised  S-procedure lossless theorem is of independent interest.  Some final remarks are described in Section
\ref{sec:5}.

\subsection* {Notation and Preliminaries} \label{sec:2}
Let $\mathbb{R}$  denote the set of real numbers. For a vector $v$, its Euclidean norm is denoted
by $|v|$.  Given a matrix $M$, the transpose and conjugate transpose are denoted respectively as $M^{T}$ and $M^{*}$.  We use  $\RE \left\lbrace \lambda\right\rbrace$
to denote the real part of a complex number $\lambda$.

Define $\Lone(-\infty,\infty) := \big\{ z : \Real \to \Real \;|\; \|z\|_1 := \int_{-\infty}^{\infty} \allowbreak  |z(t)| dt < \infty \big\}$.
We use $\Lone^{+}(-\infty,\infty)$ to denote the set of all $z(t)\in \Lone(-\infty,\infty)$ satisfying $z(t)\ge 0,\forall t$. Given a signal $z(t)\in \Lone(-\infty,\infty)$,
denote by $Z(j\omega)$ its Fourier transform and $\mathbf{Z}$ the convolution operator whose kernel is $z(t)$.  Let $ \LT(-\infty,\infty) := \left\lbrace x: \Real \to \mathbb{R}^n \;|\; \|x\|^2 := \int_{-\infty}^{\infty} |x(t)|^2 dt < \infty \right\rbrace $ and
$ \LT[0,\infty)\allowbreak := \left\lbrace x \in \LT(-\infty, \infty) \;|\; x(t) = 0 \text{ for } t < 0 \right\rbrace $.  Given
$x(t),\allowbreak y(t)\in \LT(-\infty,\infty)$, their inner product is given by $\left\langle x,y\right\rangle := \int_{-\infty}^{\infty} x(t)y(t)dt$. A bounded linear operator $\Pi$ mapping $\LT(-\infty,\infty)$ into $\LT(-\infty,\infty)$ is said to be self-adjoint if $\langle x, \Pi y \rangle = \langle \Pi x, y \rangle$ for all $x, y \in \LT(-\infty,\infty)$. A self-adjoint $\Pi$ is said to be positive if $\langle x, \Pi x \rangle \geq 0$ for all $x \in \LT(-\infty, \infty)$. The Fourier transform of
$x(t) \in \LT(-\infty, \infty)$ is denoted as $\hat{x}(j\omega)$. For any $x : \Real \to \Real^n$, define the truncation operator $(P_Tx)(t):=x(t)$ for $t\in (\infty,T]$ and $(P_Tx)(t):=0$ for
$t>T$, and the extended $\LT$-space
$ \LTE [0,\infty):= \left\lbrace x: \Real \to \mathbb{R}^n \;|\; P_T x\in \LT [0,\infty) \; \forall T\in [0,\infty) \right\rbrace$.

An operator $H :\LTE[0,\infty) \to \LTE[0,\infty)$ is said to be causal if $P_T H P_T=P_T H$ for all $T>0$ and anti-causal if
$(I - P_T) H (I - P_T) = (I - P_T)H$ for all $T > 0$. It is said to be static (a.k.a. memoryless) if it is simultaneously causal and anticausal. Let the shift operator $S_\tau : \LT(-\infty, \infty) \to \LT(-\infty, \infty)$ be defined by $(S_\tau f)(t) = f(t - \tau)$ for $\tau \in \Real$.  An operator $H$ is said to be time-invariant if $HS_{\tau}=S_{\tau}H$ for all $\tau\in\Real$. A causal operator $H :\LTE[0,\infty) \to \LTE[0,\infty)$ is said to be bounded if
\[
\|H\|:= \sup_{T>0;\|P_T u\|\neq 0} \frac{\|P_T Hu\|}{\|P_T u\|}=\sup_{0\neq u\in \LT} \frac{\|Hu\|}{\|u\|}<\infty.
\]
For a single-input single-ouput  system $H$, note that $H$ is static and time-invariant if there exists $N:\mathbb{R} \to \mathbb{R}$ such that $(H v)(t)=N(v(t))$ for all $t\in[0,\infty)$. In addition, $H$
is said to be monotone if $N$ is monotone, i.e. $x_1\ge x_2$ implies $N(x_1)\ge N(x_2)$. It is odd if $N(-x) = -N(x)$ $\forall x\in\mathbb{R}$. For notational convenience, we do not distinguish between $H$ and $N$ when $H$ is static and time-invariant.

Denote by $\mathcal{L}_\infty$ the set of transfer functions that are essentially bounded on the imaginary axis.
Every element in $\mathcal{L}_\infty$ may be associated with a bounded LTI operator  $H$   mapping from $\LT(-\infty, \infty)$ into $\LT(-\infty, \infty)$, and its induced norm is denoted by 
\[\|H\|:= \underset{u\in\mathcal{L}_2(-\infty,\infty)}{\sup}\frac{\|Hu\|}{\|u\|}.
\]
 Denote $\mathcal{RH}_{\infty}$ as the space of proper real-rational transfer functions with no poles in the closed right half
plane. Every element $G \in \mathcal{RH}_{\infty}$ is associated with a causal bounded LTI operator $G : \LTE[0,\infty) \to \LTE[0,\infty)$, which we
do not differentiate for notational convenience \cite{dullerud2013course}.

The main object of study in this work is the feedback interconnection of a $G \in \mathcal{RH}_{\infty}$ and a causal bounded
$\Delta:\LTE \to \LTE$, as illustrated in Figure \ref{fig:1}. Denote the feedback system as $[G,\Delta]$.

\begin{defn}
  $[G,\Delta]$ is said to be well-posed if the map
  $\left[ \begin{array}{c}u_1\\u_2\end{array} \right] \mapsto \left[ \begin{array}{c}d_1\\d_2 \end{array} \right]$ in Figure \ref{fig:1} has a
  causal inverse on $\LTE[0,\infty)$. It is said to be stable if it is well-posed and the inverse is bounded, in which case $[G, \Delta]$ is also used
  to denote the map $\left[ \begin{array}{c}d_1\\d_2\end{array} \right] \mapsto \left[ \begin{array}{c}u_1\\u_2 \end{array} \right]$.
\end{defn}
For the feedback system $\left[G,\Delta \right]$ as shown in Figure \ref{fig:1}, its norm is defined as 
\[
\| \left[G,\Delta \right] \|:= \sup_{T>0;\|P_T d\|\neq 0} \frac{\|P_T u\|}{\|P_T d\|},
\]
where $d:=\left[\begin{array}{c}
d_1\\d_2
\end{array}\right]$ and $u:=\left[\begin{array}{c}
u_1\\u_2
\end{array}\right]$.

We define uniform feedback stability as follows.
\begin{defn}
  The feedback system $[G,\Delta]$ is said to be uniformly stable over $\mathbf{\Delta}$ if $[G,\Delta]$ is stable for all
  $\Delta\in \mathbf{\Delta}$, and there exists $\gamma$ such that
  \[
  \sup_{\Delta\in \mathbf{\Delta}} \|[G,\Delta]\|<\gamma.
  \]
\end{defn}
This work is concerned with the uniform stability analysis of the interconnection system shown in Figure \ref{fig:1} over different classes of $\Delta$'s
using the Zames-Falb multipliers.
\begin{defn}
  A pair $v, w \in \Ltwo[0, \infty)$ is said to satisfy the IQC defined by $\Pi \in \mathcal{L}_\infty$ satisfying $\Pi(j\omega)^* = \Pi(j\omega)$ for
  all $\omega \in \mathbb{R}$ if
\begin{align} \label{eq: IQC}
\sigma_\Pi(v, w) = \int_{-\infty}^\infty \TwoOne{\hat{v}(j\omega)}{\hat{w}(j\omega)}^* \Pi(j\omega) \TwoOne{\hat{v}(j\omega)}{\hat{w}(j\omega)} d\omega \geq 0.
\end{align}
A bounded causal system $\Delta$ is said to satisfy the IQC defined by $\Pi \in \mathcal{L}_\infty$, denoted by $\Delta \in \text{IQC} (\Pi)$, if
\eqref{eq: IQC} holds for all $v \in \Ltwo[0, \infty)$ and $w = \Delta v$.
\end{defn}

\section{S-procedure lossless theorem}\label{sec: s-lemma}

In this section, the S-procedure lossless theorem in~\cite[Theorem 3.1]{MegTre93} (see also~\cite[Theorem 7]{Jonsson01lecturenotes}) is first generalised
to involve a countably infinite number of quadratic forms. This is then applied to generalising~\cite[Proposition 6]{Jonsson01lecturenotes} under a
countably infinite number of IQCs.

Denote by $\mathbb{Z}_0^+$ and $\mathbb{Z}^+$ the sets of nonnegative integers and positive integers, respectively. Define 
\[
\ell_\infty := \left\{v : \mathbb{Z}_0^+ \to \mathbb{R} : \|v\|_\infty := \sup_{i \in \mathbb{Z}_0^+} |v_i| < \infty \right\}.
\]
Let $(X, \Sigma, \mu)$ be a measure space and $L_\infty(X, \Sigma, \mu)$ denote the Banach space of essentially bounded measurable functions equipped with the norm
\[
\|u\|_\infty = \inf \{c > 0 : |u(x)| \leq c \ \mu \text{-almost everywhere}\}.
\]
Note that one may identify $\ell_\infty$ with $L_\infty(\mathbb{Z}_0^+, P(\mathbb{Z}_0^+), \lambda)$, where $P(S)$ denotes the power set of $S$ and $\lambda$ is the counting measure satisfying $\lambda(S) =$ the cardinality of $S$ if $S$ is finite and $\lambda(S) = \infty$ otherwise. Obviously, $\lambda(S) = 0$ if and only if $S = \emptyset$. For notational convenience, we write $\ell_\infty = L_\infty(\mathbb{Z}_0^+, P(\mathbb{Z}_0^+), \lambda)$, i.e. we do not differentiate between sequences in $\ell_\infty$ and functions in $L_\infty(\mathbb{Z}_0^+, P(\mathbb{Z}_0^+), \lambda)$.

A bounded, finitely additive signed measure on $P(\mathbb{Z}_0^+)$ is a signed measure $\nu : P(\mathbb{Z}_0^+) \to \Real$ satisfying $\nu(\emptyset) = 0$, $\displaystyle\sup_{A \in P(\mathbb{Z}_0^+)} |\nu(A)| < \infty$, and 
\[
\nu(A \cup B) = \nu(A) + \nu(B)
\]
for all $A, B \in P(\mathbb{Z}_0^+)$ such that $A \cap B = \emptyset$. Denote by $\mathrm{ba}(P(\mathbb{Z}_0^+))$ the set of bounded, finitely additive signed measures on $P(\mathbb{Z}_0^+)$. The dual space of $\ell_\infty = L_\infty(\mathbb{Z}_0^+, P(\mathbb{Z}_0^+), \lambda)$ is $ba(P(\mathbb{Z}_0^+)$~\cite[Theorem 3.1]{toland2020dual}, i.e. for every bounded linear functional $g$ on $\ell_\infty$ there exists $\nu \in \mathrm{ba}(P(\mathbb{Z}_0^+))$ such that 
\[
g(u) = \int_{\mathbb{Z}_0^+} u \, d\nu \quad \forall u \in \ell_\infty.
\]

Define the quadratic forms $\sigma_k : \LT[0,\infty) \to \Real$ as
\[
\sigma_k(f) = \langle f, \Pi_k f \rangle, \qquad k = 0, 1, \ldots,
\]
where $\Pi_k : \LT(-\infty, \infty) \to \LT(-\infty, \infty)$. Let $\mathcal{H} \subset \LT[0, \infty)$ be such that $S_\tau \mathcal{H} \subset \mathcal{H}$ for all $\tau > 0$. 

\begin{assum} \label{as: Pi}
Assume that 
\begin{itemize}
    \item $\Pi_k : \LT(-\infty, \infty) \to \LT(-\infty, \infty)$ is bounded LTI self-adjoint for all $k \in \mathbb{Z}_0^+$;

    \item $\sup_{k} \|\Pi_k\| < \infty$;
    
     \item for any $f_1, f_2 \in \mathcal{H}$, $\sup_k |\langle \Pi_k f_1, S_{\tau} f_2 \rangle| \to 0 $ as $\tau \to \infty$;
    
    \item there exists $f^* \in \mathcal{H}$ and $\epsilon > 0$ such that $\sigma_k(f^*) \geq \epsilon$ for $k = 1, 2, \ldots$.
\end{itemize} 
\end{assum}

\begin{thm} \label{thm: S_prod}
 Suppose that Assumption~\ref{as: Pi} holds. Then the following are equivalent:
\begin{enumerate} \renewcommand{\theenumi}{\textup{(\roman{enumi})}}\renewcommand{\labelenumi}{\theenumi}
\item \label{item: S1} $\sigma_0(f) \leq 0$ for all $f \in \mathcal{H}$ such that $\sigma_k(f) \geq 0$ for all $k = 1, 2, \ldots$;

\item \label{item: S2} There exists $\nu \in \mathrm{ba}(P(\mathbb{Z}^+))$ such that $\nu(S) \geq 0$ for all $S \in P(\mathbb{Z}^+)$ and
  \[
\sigma_0(f) + \int_{\mathbb{Z}^+} \sigma(f) \, d\nu \leq 0 
\]
for all $f \in \mathcal{H}$, where $\sigma(f) := (\sigma_1(f), \ldots, \sigma_k(f), \allowbreak  \ldots)\in \ell_{\infty}$.
\end{enumerate}
\end{thm}

\begin{pf}
  That \ref{item: S2} implies \ref{item: S1} is obvious. To see that \ref{item: S1} implies \ref{item: S2}, first note that since $\sup_k\|\Pi_k\| < \infty$ by Assumption~\ref{as: Pi}, $(\sigma_0(f), \sigma_1(f), \ldots, \sigma_k(f), \ldots  ) \in \ell_\infty$ for all $f \in \LT[0, \infty)$. Define
  \begin{align*}
  \mathcal{K} & = \left\{(\sigma_0(f), \sigma_1(f), \ldots, \sigma_k(f), \ldots  ) \in \ell_\infty : f\in \mathcal{H} \right\} \\
    \mathcal{N} & = \left\{(n_0, n_1, \ldots) \in \ell_\infty : \exists \epsilon > 0 \text{ s. t. } n_k > \epsilon \; \forall k \in \mathbb{Z}_0^+ \right\}.
  \end{align*}
  We show below that $\bar{\mathcal{K}}$, the closure of $\mathcal{K}$, is convex. Let
  $f_1, f_2 \in \mathcal{H}$, 
  \begin{align*}
    k_1 &= \left(\sigma_0(f_1), \sigma_1(f_1), \ldots, \sigma_k(f_1), \ldots\right) \in \mathcal{K}  \\
    k_2 &= \left(\sigma_0(f_2), \sigma_1(f_2), \ldots, \sigma_k(f_2), \ldots \right) \in \mathcal{K}.
  \end{align*}
  Note that since every $\Pi_k$ is bounded LTI self-adjoint,
  $ (\sigma_0(S_\tau f_2), \sigma_1(S_\tau f_2), \ldots, \sigma_k(S_\tau f_2), \ldots) = k_2$ for all $\tau > 0$. Furthermore, for all $\lambda \in [0, 1]$,
  \begin{align*}
    & \sigma_k(\sqrt{\lambda} f_1 + \sqrt{1 - \lambda} S_\tau f_2) \\
    = \; & \lambda \sigma_k(f_1) + (1 - \lambda) \sigma_k(f_2) + 2\sqrt{\lambda(1-\lambda)} \left(\langle \Pi_k f_1, S_\tau f_2 \rangle\right).
  \end{align*}
Since $\sup_{k\in\mathbb{Z}^+} \Big|\langle \Pi_k f_1, S_\tau f_2 \rangle \Big| \to 0$ as $\tau \to \infty$ by Assumption \ref{as: Pi}, it follows that for all $\epsilon > 0$ there exists sufficiently large $\alpha \in \mathbb{Z}^+$ such that 
\begin{align*}
  \sup_{k\in\mathbb{Z}^+} \Big| \langle \Pi_k  f_1, S_\tau f_2 \rangle \Big| & < \epsilon,\; \forall \tau \ge \alpha.
  \end{align*} 
 In other words, for all $\epsilon > 0$, there exists $\tau \in \mathbb{Z}^+$ such that
  \begin{align*}
    & \Big\| (\sigma_0(\sqrt{\lambda} f_1 + \sqrt{1 - \lambda} S_\tau f_2), \\
    & \qquad\qquad\qquad\sigma_1(\sqrt{\lambda} f_1 + \sqrt{1 - \lambda} S_\tau f_2), \ldots \\
     & \qquad\qquad\qquad \sigma_k(\sqrt{\lambda} f_1 + \sqrt{1 - \lambda} S_\tau f_2), \ldots  ) \\
    & \qquad\qquad\qquad\qquad\qquad\qquad - (\lambda k_1 + (1 - \lambda)k_2) \Big\|_\infty < \epsilon
\end{align*}
whereby $\lambda k_1 + (1- \lambda) k_2 \in \bar{\mathcal{K}}$. That is, $\bar{\mathcal{K}}$ is convex.

Since $\mathcal{N}$ is open and \ref{item: S1} implies that $\bar{\mathcal{K}} \cap \mathcal{N} = \emptyset$, it follows from the Hahn-Banach
theorem~\cite[Theorem 1.6]{Bre11} that there exists a closed hyperplane that separates $\bar{\mathcal{K}}$ and $\mathcal{N}$. That is, there exists $\nu_0 \in \mathrm{ba}(P(\mathbb{Z}_0^+))$ such that 
\begin{align} \label{eq: sep_H}
\int_{\mathbb{Z}_0^+} n \, d\nu_0 > 0 \quad \forall n \in \mathcal{N}
\end{align}
and 
\begin{align} \label{eq: sep_K}
\int_{\mathbb{Z}_0^+} \kappa \, d\nu_0 \leq 0 \quad \forall \kappa \in \bar{\mathcal{K}}.
\end{align}
Note that
\eqref{eq: sep_H} being true for all $n \in \mathcal{N}$ implies that $\nu_0$ is a nonnegative measure, i.e., $\nu_0(S) \geq 0$ for all $S \in P(\mathbb{Z}_0^+)$. To see this, suppose for some $S \in P(\mathbb{Z}_0^+)$, $\nu_0(S) < 0$. Let $n \in \mathcal{N}$ be $d > 0$ on $S$ and $1/d$ on $\mathbb{Z}_0^+ \setminus S$. Then since $\nu_0$ is bounded,
\[
\int_{\mathbb{Z}_0^+} n \, d\nu_0 = d\nu_0(S) + \frac{1}{d}\nu_0(\mathbb{Z}_0^+ \setminus S),
\]
which is negative for sufficiently large $d$.

Finally, let $\kappa_0 = \sigma_0(f^*)$ and $\kappa_k = \sigma_k(f^*)$ for $k = 0, 1, \ldots$. Note that $\kappa_k \geq \epsilon > 0$ for $k = 1, 2, \ldots$ by Assumption~\ref{as: Pi}. It follows from
\eqref{eq: sep_K} that
\[
\int_{\mathbb{Z}_0^+} \kappa \, d\nu_0 = \nu_0(\{0\}) \kappa_0 + \int_{\mathbb{Z}^+} (\kappa_1, \kappa_2, \ldots) \, d\nu_0\leq 0.
\]
Suppose to the contrapositive that $\nu_0(\{0\}) = 0$, then $\int_{\mathbb{Z}^+} (\kappa_1, \kappa_2, \ldots) \, d\nu_0 = 0$. Letting $n := (1, \kappa_1, \kappa_2, \ldots)$ then yields $\int_{\mathbb{Z}_0^+} n \, d\nu_0 = 0$, which violates \eqref{eq: sep_H}. Hence, $\nu_0(\{0\}) > 0$. Define $\nu \in \mathrm{ba}(P(\mathbb{Z}^+))$ as 
\[
\nu(S) := \frac{\nu_0(S)}{\nu_0(\{0\})}
\]
for all $S \in P(\mathbb{Z}^+)$. Dividing \eqref{eq: sep_K} by $\nu_0(\{0\})$ and substituting $\nu$ into it then yields \ref{item: S2}, as required. $\hfill$\qed
\end{pf}

The necessity condition in Theorem~\ref{thm: S_prod} may be simplified when the multipliers $\Pi_k, k \in \mathbb{Z}^+$, belong to a certain convex cone.

\begin{assum} \label{as: cone}
Let $\mathcal{C} \subset \mathcal{L}_{\infty}$ have the following properties:
\begin{itemize}
    \item $\mathcal{C}$ takes the form
    \[
    \mathcal{C} = \left\{\TwoTwo{0}{M}{M^*}{0} : M \in \mathcal{M} \right\};
    \]
    
    \item there exists $\mathcal{S} \subset \LT(-\infty, \infty)$ such that $M \in \mathcal{M}$ if and only if $\langle f_1, M f_2 \rangle \geq 0$ for all $(f_1, f_2) \in \mathcal{S}$.
\end{itemize}
\end{assum}

\begin{lem} \label{lem: S_nec}
 Suppose that Assumption~\ref{as: Pi} holds and let $\mathcal{C}\subset \mathcal{L}_{\infty}$ satisfy Assumption~\ref{as: cone}.
If $\Pi_k \in \mathcal{C}$ for all $k \in \mathbb{Z}^+$, then $\sigma_0(f) \leq 0$ for all $f \in \mathcal{H}$ such that $\sigma_k(f) \geq 0$ for all $k = 1, 2, \ldots$ implies that there exists $\Pi \in \mathcal{C}$ such that
\[
\sigma_0(f) +\langle f, \Pi f \rangle \leq 0
\]
for all $f \in \mathcal{H}$. 
\end{lem}

\begin{pf}
First, by applying Theorem~\ref{thm: S_prod}, it holds that there exists $\nu \in \mathrm{ba}(P(\mathbb{Z}^+))$ such that $\nu(S) \geq 0$ for all $S \in P(\mathbb{Z}^+)$ and
\begin{align} \label{eq: sigma0}
\sigma_0(f) + \int_{\mathbb{Z}^+} \sigma(f) \, d\nu \leq 0 
\end{align}
for all $f \in \mathcal{H}$, where $\sigma(f) := (\sigma_1(f), \ldots, \sigma_k(f), \allowbreak \ldots)\in \ell_\infty$. 

 Since $\Pi_k \in \mathcal{C}$ for all $k \in \mathbb{Z}^+$, it follows from Assumption~\ref{as: cone} that there exists $M_k \in \mathcal{M}$ such that 
\[
\Pi_k = \TwoTwo{0}{M_k}{M_k^*}{0}.
\]

Now let $F : \mathbb{Z}^+ \to \mathcal{L}_{\infty}$ satisfy $F(k) = M_k$ for $k \in \mathbb{Z}^+$. 
Given $f_2 \in \mathcal{L}_2(-\infty,\infty)$, set
\[
H(f_1) := \int_{\mathbb{Z}^+} \langle f_1, F(\cdot) f_2 \rangle \, d\nu = \int_{\mathbb{Z}^+} s(f_1, f_2) \, d\nu,
\]
where 
\begin{align*}
& s(f_1, f_2) \\ 
:= \, & \left(\langle f_1, M_1 f_2 \rangle, \ldots, \langle f_1, M_k f_2 \rangle, \ldots \right) \in \ell_\infty.
\end{align*}
Let $\alpha := \sup_k\|\Pi_k\|$ and since $\nu$ defines a bounded linear functional on $\ell_\infty$ with bound $\gamma > 0$, we have
\[
|H(f_1)| \leq \int_{\mathbb{Z}^+} |\langle f_1, F(\cdot) f_2 \rangle| \, d\nu \leq \alpha \gamma \|f_2\| \|f_1\|.
\]
That is, $H$ is bounded. Applying the Riesz representation theorem~\cite[Theorem A.3.52]{curtain2012introduction} yields the existence of a unique $f_F(f_2) \in \mathcal{L}_2(-\infty,\infty)$ such that
\[
\langle f_1, f_F(f_2) \rangle = H(f_1) = \int_{\mathbb{Z}^+} \langle f_1, F(\cdot) f_2 \rangle \, d\nu.
\]
By repeating the arguments above, we also have that there exists a unique $f_{F^*}(f_1) \in \mathcal{L}_2(-\infty,\infty)$ such that
\[
\langle f_2, f_{F^*}(f_1) \rangle = \int_{\mathbb{Z}^+} \langle f_2, F(\cdot)^* f_1 \rangle \, d\nu.
\]
Consider now the mapping $f_2 \mapsto f_F(f_2)$. By mimicking the arguments in \cite[Lemma A.5.9]{curtain2012introduction}, we next show that $f_2 \mapsto f_F(f_2)$ is linear and bounded. First note that $f_F(f_2)$ is linear in $f_2$ from the uniqueness of $f_F(f_2)$. We next establish using the closed graph theorem that $f_2 \mapsto f_F(z_2)$ is bounded. If $f_2^n \to f_2$ and $f_F(f_2^n) \to h$ (as $n \to \infty$), then for all $f_1 \in \mathcal{L}_2(-\infty,\infty)$ 
\[
\langle f_1, f_F(f_2^n) \rangle \to \langle f_1, h \rangle
\]
and 
\begin{align*}
    \langle f_1, f_F(f_2^n) \rangle & = \int_{\mathbb{Z}^+} \langle f_1, F(\cdot) f_2^n \rangle \, d\nu \\
    & = \int_{\mathbb{Z}^+} \langle f_2^n, F(\cdot)^* f_1 \rangle \, d\nu \\
    & = \langle f_2^n, f_{F^*}(f_1) \rangle \to \langle f_2, f_{F^*}(f_1) \rangle \\
    & \qquad \qquad \qquad \quad = \int_{\mathbb{Z}^+} \langle f_2, F(\cdot)^* f_1 \rangle \, d\nu \\
    & \qquad \qquad \qquad \quad = \langle f_1, f_F(f_2) \rangle.
\end{align*}
Thus, $h = f_F(f_2)$ and $f_2 \mapsto f_F(z_2)$ has a closed graph. By the closed graph theorem~\cite[Theorem A.3.49]{curtain2012introduction}, $f_2 \mapsto f_F(z_2)$ is bounded.

Let $M := f_2 \mapsto f_F(z_2)$, we then have $M \in \mathcal{L}_{\infty}$ and 
\begin{align} \label{eq: weak_int}
\langle f_1, M f_2 \rangle = \int_{\mathbb{Z}^+} s(f_1, f_2) \, d\nu 
\end{align}
for all $f_1, f_2 \in \mathcal{L}_2(-\infty,\infty)$.
Suppose to the contrapositive that $M \notin \mathcal{M}$, then there exists, by Assumption~\ref{as: cone}, $(f_1, f_2) \in \mathcal{S}$ such that $\langle f_1, M f_2 \rangle < 0$. However, this violates \eqref{eq: weak_int} because $\langle f_1, M_k f_2 \rangle \geq 0$ for all $k \in \mathbb{Z}^+$ and $\nu$ is a nonnegative measure.
The claim of the theorem then follows by noting that 
\[
\Pi := \TwoTwo{0}{M}{M^*}{0} \in \mathcal{C}
\]
and substituting \eqref{eq: weak_int} into \eqref{eq: sigma0}. $\hfill$\qed
\end{pf}

  By exploiting Lemma~\ref{lem: S_nec} and mimicking the proof for~\cite[Proposition 6]{Jonsson01lecturenotes}, one may obtain the following result.
  
  \begin{thm} \label{thm: sta_nec} Let $\mathcal{C} \subset \mathcal{L}_\infty$ satisfy Assumption~\ref{as: cone} and
  \begin{align*}
  \mathcal{H} := \text{cl. } \Bigg\{\TwoOne{v}{w} \in \LT[0, \infty)\ & \Bigg|\  w = \Delta v \\ & \text{for some causal bounded } \Delta \Bigg\},
  \end{align*}
  where cl. denotes the closure.
  Given LTI bounded self-adjoint $\Pi_k \in \mathcal{C}$ for $k = 1, 2, \ldots$, let
    \begin{align*}
      \mathbf{S} = \{& \Delta : \LTE[0,\infty) \to \LTE[0,\infty)\ |\ \Delta \text{ is causal,} \\ 
      & \text{bounded, and } \Delta \in \mathrm{IQC}\left(\Pi_k\right), k = 1, 2,
      \ldots\}.
    \end{align*}
    Suppose $\sup_k\|\Pi_k\| < \infty$, $\sup_k |\langle \Pi_k f_1, S_\tau f_2 \rangle | \to 0$ as $\tau \to \infty$ for all $f_1, f_2 \in \mathcal{H}$, and there exist $(v^*, w^*) \in \mathcal{H}$, $\epsilon > 0$ such that $\sigma_{\Pi_k}(v^*, w^*) \geq \epsilon$ for all $k = 1, 2, \ldots$.
    Suppose $[G,\Delta]$ is well-posed for
    all $\Delta \in \mathbf{S}$, then $[G,\Delta]$ is uniformly stable over $\mathbf{S}$ only if there exists $\Pi \in \mathcal{C}$ such that for all $\omega \in \mathbb{R}$,
\begin{align} \label{eq: S_nec}
\TwoOne{G(j\omega)}{1}^* \Pi(j\omega) \TwoOne{G(j\omega)}{1} \le -1.
\end{align}
\end{thm}

\begin{pf}
$[G,\Delta]$ in Figure \ref{fig:1} is uniformly stable over $\mathbf{S}$ implies that when $d_1=0$ (i.e., $u_1=y_2$),  there exists $\gamma>0$ such that 
\[
\sup_{T>0;\|P_T d\|\neq 0} \frac{\|P_T y_2\|}{\|P_T d_2\|}<\gamma.
\]
Let 
 \begin{align*}
      \sigma_0(u_2, y_2) & := \|y_2\|^2 +\|Gy_2\|^2- \gamma \|u_2 - Gy_2\|^2; \\
      \sigma_k(u_2, y_2) & := \sigma_{\Pi_k}(u_2, y_2).
  \end{align*}
The uniform stability of $[G,\Delta]$ over $\mathbf{S}$ means that $\sigma_0(u_2, y_2) \leq 0$ for all $(u_2, y_2) \in \mathcal{H}$ such that $\sigma_k(u_2, y_2) \geq 0$, $k = 1, 2, \ldots$. By Lemma~\ref{lem: S_nec}, this implies that there exists $\Pi \in \mathcal{C}$ such that
  \[
  \sigma_0(u_2, y_2) + \left\langle \TwoOne{u_2}{y_2}, \Pi \TwoOne{u_2}{y_2} \right\rangle \leq 0
  \]
  for all $(u_2, y_2) \in \mathcal{H}$. Setting $u_2 = Gy_2$ and applying~\cite[Proposition 4]{Jonsson01lecturenotes} then yields \eqref{eq: S_nec}. In particular, note that since $G \in \mathcal{R}\mathcal{H}_\infty$, 
  \[
  \TwoOne{Gy_2}{y_2} \in \mathcal{H},
  \]
  as there always exist bounded causal (nonlinear) $\{\Delta_i\}$ for which $\|\Delta_i G y_2 - y_2\| \to 0$ given any $y_2 \in \LT[0, \infty)$. To see this, it suffices to consider $G(j\infty) \neq 0$ and let $\Delta (G y_2) := y_2$. Furthermore, for all $u_2 \in \LT[0, \infty)$ such that $u_2(t) = (G y_2)(t)$ on some interval $[0, T)$ and $u(t) \neq (G y_2)(t)$ on $[T, T + \epsilon)$ for some $\epsilon > 0$, let $(\Delta u_2) (t) := y_2(t)$ on $[0, T)$ and $(\Delta u) (t) := 0$ otherwise. Repeating the above procedure and setting to $0$ the output of $\Delta$ to all input $u_2(t) \neq G y_2(t)$ on any interval $[0, T)$, it may be verified that such a $\Delta$ is bounded and causal. $\hfill$\qed
\end{pf}
  
\section{Robust stability against monotone nonlinearity} \label{sec:3} 
Let $\mathbf{\Delta_{0}}$ consist of all causal bounded
system $\Delta$ that maps $0$ into $0$ and satisfies $\langle x(t),(\Delta x)(t)\rangle\ge 0,\forall x\in \LT[0,\infty)$.  For $\beta \in (1, \infty]$, consider the inequality
\begin{equation}\label{eq:58}
\begin{split}
\int_{0}^{\infty} x(t+\tau)(\Delta x)(t)dt \le \int_{0}^{\infty} x(t)(\Delta x)(t)dt,\\
\forall \tau\in (-\beta,-1/\beta) \cup (1/\beta,\beta), x\in \LT[0,\infty)  
\end{split}
\end{equation}
and define 
\begin{equation}\label{eq: set of delta}
\mathbf{\Delta}_\beta := \left\lbrace \Delta \in \mathbf{\Delta_0} \; |\; \Delta\text{ satisfies } \eqref{eq:58} \right\rbrace.
\end{equation}
Specifically, $\mathbf{\Delta}_\infty$ consists of $\Delta \in \mathbf{\Delta_0}$ that satisfies \eqref{eq:58} for all $\tau \in \Real$.

\subsection{Nonlinear dynamic uncertainty}
Denote by $\mathbf{Z}$ the convolution operator whose kernel is $z\in \Lone(-\infty,\infty)$. Let $\Lone(\beta)$ (respectively, $\Lone^{+}(\beta$) be the subspace of  $\Lone(-\infty,\infty)$ (respectively, $\Lone^+(-\infty,\infty)$) which consists of all $z(t)\in\Lone(-\infty,\infty)$ (respectively, $z(t)\in\Lone^+(-\infty,\infty)$) satisfying $z(t)=0,\forall t\notin \{(-\beta,-1/\beta)\cup(1/\beta,\beta)\}$.

The following lemma characterizes the input-output
relations of all elements in the set $\mathbf{\Delta}_\beta$ by a class of IQCs defined by the Zames-Falb multipliers.
\begin{lem}\label{le:1}
Given a $\Delta\in \mathbf{\Delta_0}$, $\Delta$ satisfies \eqref{eq:58} if and only if 
\begin{equation}\label{eq:62}
\left\langle x(\cdot), (1-\mathbf{Z})(\Delta x)(\cdot)\right\rangle \ge 0, \forall x\in \LT[0,\infty)
\end{equation}
for all  $z\in \Lone^+(\beta)$ such that  $\|z\|_1\le  1$.
\end{lem}
\pf
``$\Rightarrow$''  First note that  $\int_{0}^{\infty}x(t)(\Delta x) (t)dt  \allowbreak \ge 0$ for all $x\in\LT[0,\infty)$ as $\Delta\in\mathbf{\Delta}_0$.
Since $x(\cdot)$ is in $\LT[0,\infty)$, we have
\[
\left\langle x(\cdot), \mathbf{Z}(\Delta x)(\cdot)\right\rangle= \int_{-\infty}^{\infty}z(\tau)\int_{0}^{\infty}x(t)(\Delta x)(t-\tau)dtd\tau.
\]
It follows from \eqref{eq:58}, $z(t)\ge 0$, and $z(t)=0$ for all $t\notin \{(-\beta,-1/\beta)\cup(1/\beta,\beta)\}$  that 
\[
\begin{split}
        &  \int_{-\infty}^{\infty}z(\tau)\int_{0}^{\infty}x(t)(\Delta x)(t-\tau)dtd\tau \\
  \le  &  \int_{-\infty}^{\infty}z(\tau)\int_{0}^{\infty}x(t)(\Delta x)(t)dtd\tau.
\end{split}
\]
Since  $\|z\|_1\le 1$ and  $\int_{0}^{\infty}x(t)(\Delta x)(t)dt \ge 0$, it follows that
\[
\int_{-\infty}^{\infty}z(\tau)\int_{0}^{\infty}x(t)(\Delta x)(t)dtd\tau\le \int_{0}^{\infty}x(t)(\Delta x)(t)dt.
\]
Combining the two inequalities above yields
\[
\left\langle x(\cdot), \mathbf{Z}(\Delta x)(\cdot)\right\rangle \le \left\langle x(\cdot), (\Delta x)(\cdot)\right\rangle.
\]
Hence, \eqref{eq:62} is shown.

``$\Leftarrow$''   Given  $\epsilon > 0$ and $\bar{\tau} \in \{(-\beta,-1/\beta)\cup(1/ \beta,\beta)\}$, let $z(t) := 1/\epsilon$ if
$t \in [\bar{\tau} - \frac{\epsilon}{2}, \bar{\tau} + \frac{\epsilon}{2}]$ and $z(t) := 0$ otherwise, whereby $\|z\|_1 = 1$. Then inequality \eqref{eq:62} gives
\[
\int_{-\infty}^{\infty}z(\tau)\int_{0}^{\infty}x(t)(\Delta x)(t-\tau)dtd\tau \le \int_{0}^{\infty}x(t)(\Delta x)(t)dt.
\]

Since
$\int_{0}^{\infty}x(t)(\Delta x)(t-\bar{\tau})dt = \lim_{\epsilon \to 0} \int_{-\infty}^{\infty}z(\tau)\int_{0}^{\infty}x(t) \allowbreak (\Delta
x)(t-\tau)dtd\tau$ and $\bar{\tau}$ can be arbitrary number in $\{(-\beta,-1/\beta)\cup(1/ \beta,\beta)\}$, this implies that $\Delta$ satisfies \eqref{eq:58}. 
$\hfill$\qed
\endpf

Next, consider the inequality 
\begin{equation}\label{eq: 58_2}
\begin{split}
\left| \int_{0}^{\infty} x(t+\tau)(\Delta x)(t)dt\right| \le \int_{0}^{\infty} x(t)(\Delta x)(t)dt,\\
 \forall \tau\in (-\beta,-1/\beta) \cup (1/\beta,\beta),\forall x\in \LT[0,\infty) 
\end{split}
\end{equation}
and define
\begin{equation}\label{eq: set of delta_bar}
\mathbf{\bar{\Delta}}_\beta := \left\lbrace \Delta \in \mathbf{\Delta_0} \; | \; \Delta \text{ satisfies }\eqref{eq: 58_2}  \right\rbrace.
\end{equation}
Specifically, $\mathbf{\bar{\Delta}}_\infty$ consists of $\Delta \in \mathbf{\Delta_0}$ that satisfies \eqref{eq: 58_2} for all $\tau \in \Real$. Evidently $\mathbf{\bar{\Delta}}_\beta$ is a subset of $\mathbf{\Delta}_\beta$. The next lemma shows that as far as $\mathbf{\bar{\Delta}}_\beta$ is concerned, a more
restrictive necessary and sufficient condition than the one presented in Lemma \ref{le:1} can be provided.

\begin{lem}\label{le: 1_2}
Given a $\Delta \in \mathbf{\Delta_0}$, $\Delta$ satisfies \eqref{eq: 58_2} if and only if \eqref{eq:62} holds
for all  $z\in \Lone(\beta)$ such that  $\|z\|_1\le  1$.
\end{lem}
\pf
Sufficiency can be proved by  following the same argument of the proof for Lemma \ref{le:1} with $\pm z(t)$ constructed therein.

To prove necessity, observe that for any $z \in \Lone(\beta)$, 
\[
\left\langle x(\cdot), \mathbf{Z}(\Delta x)(\cdot)\right\rangle \le \int_{-\infty}^{\infty}\left|z(\tau)\right| \left| \int_{0}^{\infty}x(t)(\Delta x)(t-\tau)dt\right|d\tau.
\]
It then follows from \eqref{eq: 58_2} that
\[
\left\langle x(\cdot), \mathbf{Z}(\Delta x)(\cdot)\right\rangle \le\int_{-\infty}^{\infty}\left|z(\tau)\right| \int_{0}^{\infty}x(t)(\Delta x)(t)dtd\tau.
\]
Since $\|z\|_1\le 1$ and $\int_{0}^{\infty}x(t)(\Delta x)(t)dt\ge 0$, it follows that 
\[
\left\langle x(\cdot), \mathbf{Z}(\Delta x)(\cdot)\right\rangle \le \left\langle x(\cdot), (\Delta x)(\cdot)\right\rangle.
\]
Hence, \eqref{eq:62} is shown. $\hfill$\qed
\endpf

The main result showing the necessary and sufficient condition for the existence of an appropriate Zames-Falb multiplier is presented below.
\begin{thm}\label{th:1}
  Consider Figure \ref{fig:1} with $G\in \mathcal{RH}_{\infty}$. Suppose $[G,\Delta]$ is well-posed for all
  $\Delta \in \mathbf{\Delta}_\infty$, then $[G,\Delta]$ is uniformly stable over $\mathbf{\Delta}_\infty$ if  
\begin{equation}\label{eq:th1}
    \begin{array}{l}
\exists z\in \Lone^+(-\infty,\infty), \epsilon>0 \text{ s.t. } \|z\|_1\le  1 \text{ and }
\\ 
\RE \left\{(1-Z(j\omega))(-G(j\omega)) \right\}\ge \epsilon, \quad \forall \omega\in \Real.
    \end{array}
\end{equation} 
Moreover, $[G,\Delta]$ is uniformly stable over $\mathbf{\Delta}_\beta$ for any $\beta \in (1, \infty)$ only if \eqref{eq:th1} holds.
\end{thm}
\pf
``$\Leftarrow$'' Letting $\beta=\infty$, it is shown by the  necessity direction of  Lemma \ref{le:1} that for all $\Delta\in \mathbf{\Delta}_{\infty}$, 
\begin{equation}\label{eq:IQCre}
   \left\langle u_2(\cdot), (1-\mathbf{Z})\Delta u_2(\cdot)\right\rangle \ge 0, \forall u_2 \in \LT[0,\infty) 
\end{equation}
for all  $z\in \Lone^{+}(-\infty,\infty)$ such that  $\|z\|_1\le 1$. Denote $\Delta u_2$ as $y_2$, and note that condition \eqref{eq:IQCre} is equivalent to $(u_2,y_2)$ satisfying the IQC 
\begin{equation}\label{eq: IQC_Delta}
\int_{-\infty}^\infty\left[\begin{array}{c} \hat{u}_2 (j\omega)\\ \hat{y}_2(j\omega)
\end{array} \right]^{*} \Pi (j\omega)\left[\begin{array}{c} \hat{u}_2 (j\omega)\\ \hat{y}_2(j\omega)
\end{array} \right] d\omega \ge 0,  \forall u_2 \in \Ltwo[0, \infty)
\end{equation}
where the Zames-Falb multiplier $\Pi$ is given by 
\begin{equation} \label{eq: multiplier 1}
\Pi(j\omega)=\left[\begin{array}{cc} 0 & 1-Z(j\omega)^{*}\\  1-Z(j\omega) & 0
\end{array}
 \right].
\end{equation}
Observe that  \eqref{eq:th1} can be rewritten as 
\begin{equation} \label{eq: IQC G}
\left[\begin{array}{c} G(j\omega) \\ 1
\end{array} \right]^{*} \Pi (j\omega)\left[\begin{array}{c} G(j\omega) \\ 1
\end{array} \right]\le -2\epsilon,  \forall \omega\in \mathbb{R}.
\end{equation}
Noting that $[G,\Delta]$ is well-posed for all $\Delta \in \mathbf{\Delta}_\infty$ and $\Pi$ is anti-diagonal, the robust stability of $[G,\Delta]$ over
$\Delta \in \mathbf{\Delta}_\infty$ follows from \cite[Theorem 1]{megretski1997system}.

To show the uniform stability, define the graphs of $G$ and $\Delta$ as, respectively,
\[
\mathcal{G}(G) :=  \left\lbrace [\begin{array}{cc}
y_1 & u_1
\end{array}]^{T}\;|\;  y_1=G u_1,\forall u_1\in \LT[0,\infty)\right\rbrace
\]
and
\[
\mathcal{G}(\Delta) :=  \left\lbrace [\begin{array}{cc}
u_2 & y_2
\end{array}]^{T} \;|\; y_2=\Delta u_2, \forall u_2\in \LT[0,\infty) \right\rbrace.
\]
Since $G\in \mathcal{RH}_\infty$, it follows that $\left\langle\nu_2,\Pi\nu_2\right\rangle\ge 0,\forall \nu_2 \in \mathcal{G}(\Delta)$ and
$\left\langle \nu_1,\Pi\nu_1 \right\rangle\le -\hat{\epsilon}\|\nu_1\|^2,\forall \nu_1 \in\mathcal{G}(G)$ for some $\hat{\epsilon} > 0$.  Summing
them gives
\[
\begin{array}{rcl}
\hat{\epsilon}\|\nu_1\|^2 & \le & \left\langle\nu_2,\Pi\nu_2\right\rangle -\left\langle \nu_1,\Pi\nu_1 \right\rangle\\
 & = & \left\langle \nu_2-\nu_1,  \Pi\nu_2-\Pi\nu_1 \right\rangle  +  \left\langle \nu_1,  \Pi\nu_2-\Pi\nu_1 \right\rangle +\\
  & &  \left\langle \nu_2-\nu_1,  \Pi\nu_1 \right\rangle\\
  & \le & \|\Pi\|\cdot\|\nu_2-\nu_1\|^2+2\|\Pi\|\cdot\|\nu_2-\nu_1\|\cdot\|\nu_1\|\\
   & \le & \|\Pi\|\cdot\|\nu_2-\nu_1\|^2+\frac{2\|\Pi\|^{2}\cdot\|\nu_2-\nu_1\|^2}{\hat{\epsilon}}+\frac{\hat{\epsilon}}{2}\|\nu_1\|^2.
\end{array}
\]
It follows that
\[
\frac{\hat{\epsilon}}{2}\|\nu_1\|^2 \le \left(  \|\Pi\| + \frac{2\|\Pi\|^{2}}{\hat{\epsilon}} \right) \|\nu_2-\nu_1\|^2.
\]
Moreover, $\|\nu_2\|^{2}=\|\nu_2-\nu_1+\nu_1\|^2\le \left( \|\nu_2-\nu_1\|+\|\nu_1\|\right)^2 \allowbreak \le 2\|\nu_2-\nu_1\|^{2}+2\|\nu_1\|^2 $. Combining the
above inequalities yields the existence of $\gamma >0$ such that
\[
 \gamma \left( \|\nu_1\|^2 +\|\nu_2\|^2 \right) \le \|\nu_2-\nu_1\|^2.
\]
Since $\nu_1\in \mathcal{G}(G)$, $\nu_2\in\mathcal{G}(\Delta)$, and $[G, \Delta]$ is robustly stable over $\Delta \in \mathbf{\Delta}_\infty$, the above
inequality leads to
\begin{align*}
\gamma \left( \|y_1\|^2 +\|u_1\|^2+\|u_2\|^2+\|y_2\|^2 \right)   \le   \|d_1\|^2+\|d_2\|^2\\
\forall d_1, d_2  \in  \Ltwo[0, \infty),\;  \forall \Delta \in \mathbf{\Delta}_\infty. \qquad  \qquad
\end{align*}
Hence, it can be concluded that $[G,\Delta]$ is uniformly stable over $\mathbf{\Delta}_\infty$.

``$\Rightarrow$'' By defining $v(t) = w(t) = 1$ for $t \in [0, 1]$ and $v(t) = w(t) = 0$ otherwise, we first show that there exists $\epsilon>0$ such that 
 $\left\langle v(\cdot), (1-\mathbf{Z})w(\cdot)\right\rangle > \epsilon$ for all $z \in \Lone(\beta)$ satisfying   $\|z\|_1 \leq 1$. 
 Observe that $\left\langle v(\cdot), w(\cdot)\right\rangle =1$ and $\left\langle v(\cdot), \mathbf{Z}w(\cdot)\right\rangle =\int_{-\infty}^{\infty}z(\tau)\int_0^{\infty} v(t)w(t-\tau) dtd\tau =\int_{-1}^{1}z(\tau)(1-|\tau|) d\tau $. Since $z(t)=0$ for all $t\notin \{(-\beta,-1/\beta)\cup(1/\beta,\beta)\}$, it follows that $\int_{-1}^{1}z(\tau)(1-|\tau|) d\tau=\int_{-1}^{-1/\beta }z(\tau)(1-|\tau|) d\tau+\int_{1/\beta}^{1}z(\tau)(1-|\tau|) d\tau\le (1-1/\beta) \int_{-1}^{1}z(\tau) d\tau$. Together with $\|z\|_1\le 1$, we have $\int_{-1}^{1}z(\tau)(1-|\tau|) d\tau \le 1-1/\beta$. Therefore, we obtain that $\left\langle v(\cdot), (1-\mathbf{Z})w(\cdot)\right\rangle \ge 1/\beta>0$. 
 
Let $z_1=0$ and
$\{z_k\}_{k = 2}^\infty$ be a normalized Schauder basis for $\Lone(\beta)$ consisting of non-negative functions \cite{johnson2015schauder}. That is, $\|z_k\|_1 = 1,k=2,3,\ldots$ and for all $z \in \Lone^+(\beta)$, there exist $\tau_k \geq 0,\; k=2,3,\ldots$ for which
$z = \sum_{k = 1}^\infty \tau_i z_i$. Define
\[
\Pi_k(j\omega) := \TwoTwo{0}{1-Z_k(j\omega)^*}{1-Z_k(j\omega)}{0},k\in \mathbb{Z}^+.
\]

Note that $\Pi_k,k\in\mathbb{Z}^+$  is bounded LTI and self-adjoint. It is  clear that $\sup_{k\in\mathbb{Z}^+} \|\Pi_k\| < \infty$. Moreover,  for any $f_1, f_2 \in \LT[0,\infty)$, $\sup_{k\in\mathbb{Z}^+} |\langle \Pi_k f_1, S_{\tau} f_2 \rangle| \to 0 $ as $\tau \to \infty$ since $\beta < \infty$. Hence, Assumption \ref{as: Pi} holds for $\Pi_k,k\in \mathbb{Z}^+$.

By Lemma \ref{le:1}, every element in $\mathbf{\Delta}_\beta$ satisfies the IQCs defined by $\Pi_k$'s for all $k\in\mathbb{Z}^+$. Let $\mathcal{C} \subset \mathcal{L}_\infty$ be of the form: 
\begin{align*}
\mathcal{C} := & \Bigg\{\alpha\TwoTwo{0}{1-Z(j\omega)^*}{1-Z(j\omega)}{0} :\alpha \geq 0, \\ & \qquad \qquad  z\in \Lone^{+}(-\infty,\infty), \|z\|_1 \leq 1\Bigg\}.
\end{align*}

Evidently, $\Pi_k\in \mathcal{C},\forall k\in\mathbb{Z}^+$. By~\cite[Theorem 1]{KulSaf02}, $\langle v, M w \rangle \geq 0$ for all $(v, w)$ in $\{(v, w) \in \LT(-\infty, \infty): w = \Delta v, \; \Delta \text{ is static, time-}  \allowbreak \text{invariant, and monotone} \}$
if and only if $M = \alpha(1-\mathbf{Z})$ for some $\alpha \geq 0$ and $z\in \Lone^{+}(-\infty,\infty)$ with $\|z\|_1 \leq 1$. That is, Assumption~\ref{as: cone} is satisfied.

The uniform feedback stability of $[G, \Delta]$ over all $\Delta\in \mathbf{\Delta}_0$ satisfying the IQCs defined by $\Pi_k$ for $k = 1, 2, \ldots$ implies 
by Theorem~\ref{thm: sta_nec} that there exists $\Pi \in \mathcal{C}$ such that for all $\omega \in \mathbb{R}$,
\[
\TwoOne{G(j\omega)}{1}^* \Pi(j\omega) \TwoOne{G(j\omega)}{1} \leq -1.
\]
This is equivalent to \eqref{eq:th1}. $\hfill$\qed \endpf
\begin{thm}
\label{th:1_2}
 Consider Figure \ref{fig:1} with $G\in \mathcal{RH}_{\infty}$. Suppose $[G,\Delta]$ is well-posed for all
  $\Delta \in \bar{\mathbf{\Delta}}_\infty$, then $[G,\Delta]$ is uniformly stable over $\bar{\mathbf{\Delta}}_\infty$ if  
  \begin{equation}\label{eq:th2}
    \begin{array}{l}
\exists z\in \Lone(-\infty,\infty), \epsilon>0 \text{ s.t. } \|z\|_1\le  1 \text{ and }
\\ 
\RE \left\{(1-Z(j\omega))(-G(j\omega)) \right\}\ge \epsilon, \quad \forall \omega\in \Real.
    \end{array}
\end{equation} 
Moreover, $[G,\Delta]$ is uniformly stable over $\bar{\mathbf{\Delta}}_\beta$ for any $\beta \in (1, \infty)$ only if \eqref{eq:th2} holds.
\end{thm}
\pf
``$\Leftarrow$''   Sufficiency follows by similar arguments in the sufficiency proof of Theorem \ref{th:1}.
 
``$\Rightarrow$''  Let $z_1=0$ and $\{z_i\}_{i = 2}^\infty$ be such that $\|z_i\|_1 = 1$ and for all $z \in \Lone(\beta)$, there exist $\tau_i \geq 0$ for
which $z = \sum_{i = 1}^\infty \tau_i z_i$~\cite{johnson2015schauder}. This can be obtained by taking a union of a normalized Schauder basis for $\Lone(\beta)$ consisting of non-negative functions and the negatives of its elements, with the functions with flipped signs ordered next to each other. One can then
readily establish necessity by applying Lemma \ref{le: 1_2} and the same arguments as those in the necessity proof of Theorem \ref{th:1}. $\hfill$\qed
\endpf

\subsection{Static uncertainty}
Next, we consider the class of static (i.e. memoryless) $\Delta$'s in the set $\mathbf{\Delta}_\infty$. The following lemma explains the links of the set
$\mathbf{\Delta}_\infty$ to the type of static nonlinearity that is frequently encountered in the control literature.
\begin{lem} \label{le: static_1}
Given a static and time-invariant $\Delta \in \mathbf{\Delta_0}$,  $\Delta\in\mathbf{\Delta}_\infty$ if and only if  $\Delta$ is monotone nondecreasing.
\end{lem}
\pf
``$\Leftarrow$'' Since $\Delta$ is assumed to be static time-invariant and bounded, there exists a constant $C>0$ such that for all $x\in\mathbb{R}$, $ |\Delta(x)|\le Cx$.  Consequently, sufficiency follows \cite[Lemma 8]{zames1967stability}. 

``$\Rightarrow$'' We prove necessity by contraposition.  Given a static and time-invariant $\Delta \in \mathbf{\Delta_0}$ that satisfies \eqref{eq:58}, suppose to the
  contrapositive that there exist $x_1,x_2 \in \mathbb{R}$ such that $x_1< x_2$ and $\Delta(x_1) > \Delta(x_2)$, i.e.  $\Delta$ is not monotone nondecreasing. It can
  be easily verified that
\[
x_2\Delta(x_1)+x_1\Delta (x_2)>x_1\Delta(x_1)+x_2\Delta(x_2).
\]
Define $x(t) \in \mathcal{L}_{2} [0,\infty)$ as 
\begin{equation}\label{eq:nec_ex}
x(t)=\left\lbrace \begin{array}{ll} x_1 & \text{if} \;  t\in [0,1)  \cup \cdots \cup [2L,2L+1)\\
 x_2 & \text{if} \;  t\in [1,2)  \cup  \cdots \cup [2L+1,2L+2)
 \\
 0 & \text{otherwise}
\end{array} \right.
\end{equation}
for some $L\in \mathbb{Z}^+$. 
Then it follows that 
\[
\begin{array}{rl}
 & x(t)\Delta(x(t+1)) - x(t)\Delta(x(t)) \\
 = & \left\lbrace
 \begin{array}{ll} x_1\Delta(x_2)-x_1\Delta(x_1) & \text{if} \;  t\in [0,1) \cup\cdots \cup [2L,2L+1)\\
 x_2 \Delta(x_1)-x_2\Delta (x_2)& \text{if} \;t\in [1,2)  \cup \cdots \cup [2L-1,2L)\\
 -x_2\Delta(x_2) &   \text{if} \; t\in [2L+1,2L+2)\\
 0 & \text{otherwise}.
\end{array} \right.
\end{array}
\]
When $L$ is chosen to be  sufficiently large,  we have that  
\[
\int_{0}^{\infty} x(t)\Delta(x(t+1))dt>\int_{0}^{\infty} x(t)\Delta(x(t))dt.
\]
Hence, from \eqref{eq: set of delta} we have  $\Delta \notin \mathbf{\Delta}_\infty$, which completes the proof.  $\hfill$\qed
\endpf

Lemmas~\ref{le:1} and \ref{le: static_1} demonstrate that the Zames-Falb IQCs used in this paper completely characterize the
  `monotonicity' property in the set of monotonic static time-invariant nonlinearities. Whereas there exist richer classes of IQCs, defined by
  multipliers consisting of convolutional impulses as in~\cite{zames1967stability} or linear time-varying multipliers~\cite{KulSaf02}, that such
  nonlinearities satisfy, these IQCs do not provide additional benefits from a theoretical perspective in that they do not capture any extra
  structures in the uncertainty. We caution that for computational or mathematical convenience, such IQCs may still be used in practice.

\begin{lem} \label{le: static_2}
Given a static and time-invariant $\Delta \in \mathbf{\Delta_0}$,  $\Delta\in\bar{\mathbf{\Delta}}_\infty$  if and only if  $\Delta$ is odd almost everywhere (a.e.)  and monotone nondecreasing.
\end{lem}
\pf
``$\Leftarrow$''  The sufficiency can be proved by a similar argument in \cite[Lemma 8]{zames1967stability}.

``$\Rightarrow$'' Given a static and time-invariant $\Delta \in \mathbf{\Delta_0}$ that belongs to $\bar{\mathbf{\Delta}}_\infty$, obviously $\Delta$ belongs  also to  $\mathbf{\Delta}_\infty$. According to Lemma \ref{le: static_1}, one has that $\Delta$ is monotone nondecreasing. Moreover, since a monotone function can only
have countable jump discontinuities, $\Delta$ is continuous a.e..  What remains to be shown is that when $\Delta$ is static,
$\Delta \in \mathbf{\bar{\Delta}}_\infty$ only if it is odd a.e.. To this end, let $x_1,x_2 $ be any scalars in $\mathbb{R}$ such that $\Delta$ is continuous
at $x_1$ and $x_2>x_1>0$. Since $\Delta(0)=0$, it follows from Lemma \ref{le: static_1} that $\Delta(x_2) \ge \Delta(x_1)\ge 0$ and
$\Delta(-x_2) \le \Delta(-x_1) \le 0$.

Now construct a signal $x(t) \in \mathcal{L}_{2} [0,\infty)$ as 
\begin{equation}\label{eq: construct x_1}
x(t)=\left\lbrace \begin{array}{ll} x_1 & \text{if} \;  t\in [0,1)  \cup \cdots \cup [2L,2L+1)\\
- x_2 & \text{if} \; t\in [1,2)  \cup  \cdots \cup [2L+1,2L+2)\\
 0 & \text{otherwise}
\end{array} \right.
\end{equation}
for some $L\in\mathbb{Z}^+$.
Then it follows that 
\begin{equation}\label{eq: construct x}
\begin{array}{l}
 x(t)\Delta(x(t+1)) + x(t)\Delta(x(t))=\\
  \left\lbrace \begin{array}{ll} x_1\Delta(-x_2)+x_1\Delta(x_1) & \text{if} \;  t\in [0,1)\cup \cdots \cup [2L,2L+1) \\
 -x_2 \Delta(x_1)-x_2\Delta (-x_2)& \text{if} \; t\in [1,2)  \cup \cdots \cup [2L-1,2L)\\
  -x_2\Delta(-x_2) &   \text{if} \; t\in [2L+1,2L+2)\\
 0 & \text{otherwise.}
\end{array} \right.
\end{array}
\end{equation}
Since $x(t)\Delta(x(t+1))\le 0$ for all $t\ge 0$, it can be inferred from $\Delta\in\mathbf{\bar{\Delta}}_\infty$ that 
\[
\int_{0}^{\infty} x(t)\Delta(x(t+1))dt\ge-\int_{0}^{\infty} x(t)\Delta(x(t))dt,
\]
which by \eqref{eq: construct x} with sufficiently large $L$ implies that
\[
\begin{array}{rl}
 &x_1\Delta(-x_2)+x_1\Delta(x_1)  -x_2 \Delta(x_1)-x_2\Delta (-x_2)\\
 =& (x_2-x_1)(-\Delta(x_1)-\Delta(-x_2))
 \ge 0.
\end{array}
\]
Hence, one has $\Delta(-x_2)\le-\Delta(x_1)$.

If we construct the signal $x(t)$ differently as
\begin{equation}\label{eq: construct x_2}
x(t)=\left\lbrace \begin{array}{ll} x_2 & \text{if} \;  t\in [0,1) \cup \cdots \cup [2L,2L+1)\\
- x_1 & \text{if} \; t\in [1,2)  \cup  \cdots \cup [2L+1,2L+2)\\
 0 & \text{otherwise}
\end{array} \right.
\end{equation}
then by a similar deduction, it can be verified that $\Delta(x_2)\ge-\Delta(-x_1)$.  Since $\Delta$ is continuous at $x_1$, taking the limit
$x_2\rightarrow x_1^{+}$ yields $\Delta(-x_2)\rightarrow \Delta(-x_1)$ and $\Delta(x_2)\rightarrow\Delta(x_1)$. Together with the two inequalities
obtained, it can be concluded that $\Delta(x_1)=-\Delta(-x_1)$, as required.
$\hfill$
\qed
\endpf
 Now let us define
\[
\begin{array}{l}
\mathbf{\Delta_{static}} :=\\
   \left\lbrace \Delta \in \mathbf{\Delta_0}  \; | \;  \Delta \text{ is a  monotone nondecreasing function} \right\rbrace.
\end{array}
\]
It follows from Lemma \ref{le: static_1} that
$\mathbf{\Delta_{static}} = \big\{ \Delta \in \mathbf{\Delta}_\infty \; | \; \Delta \allowbreak \text{ is static and time-invariant} \big\}$. Therefore, it is clear that
$\mathbf{\Delta_{static}}\subset\mathbf{\Delta}_\infty $, and the following result follows directly from Theorem \ref{th:1}.
\begin{cor}
  Consider Figure \ref{fig:1} where $G\in \mathcal{RH}_{\infty}$ and $\Delta \in \mathbf{\Delta_{static}}$, and suppose $[G,\Delta]$ is well-posed for
  all $\Delta \in \mathbf{\Delta_{static}}$.  Then, $[G,\Delta]$ is uniformly stable over $ \mathbf{\Delta_{static}}$ if  \eqref{eq:th1} holds.
\end{cor}
Analogously, define 
\[
\mathbf{\bar{\Delta}_{static}} := \left\lbrace \Delta \in \mathbf{\Delta_{static}}  \; | \;  \Delta  \text{ is odd a.e.}   \right\rbrace.
\]
By Lemma \ref{le: static_2},
\[\mathbf{\bar{\Delta}_{static}} = \left\lbrace \Delta \in \mathbf{\bar{\Delta}}_\infty \; | \; \Delta \text{ is static and time-invariant} \right\rbrace.\] 
Hence, the following
result follows from Theorem \ref{th:1_2} as $\mathbf{\bar{\Delta}_{static}}\subset\mathbf{\bar{\Delta}}_\infty$.
\begin{cor}\label{cor: static_2}
  Consider Figure \ref{fig:1} where $G\in \mathcal{RH}_{\infty}$ and $\Delta \in \mathbf{\bar{\Delta}_{static}}$, and suppose $[G,\Delta]$ is
  well-posed for all $\Delta \in \mathbf{\bar{\Delta}_{static}}$.  Then, $[G,\Delta]$ is uniformly stable over $ \mathbf{\bar{\Delta}_{static}}$ if
   \eqref{eq:th2} holds.
\end{cor}

\begin{rem}
  The preceding corollaries derive from Theorems \ref{th:1} and \ref{th:1_2} the classical results of applying the Zames-Falb multipliers to establishing
  stability of the feedback interconnection depicted in Figure \ref{fig:1} with $G\in\mathcal{RH}_{\infty}$ and $\Delta$ being  a static  time-invariant
  monotone nonlinearity. As $\mathbf{\Delta_{static}}$ is a subset of $\mathbf{\Delta}_\beta$ (respectively,
  $\mathbf{\bar{\Delta}_{static}}\subset\mathbf{\bar{\Delta}}_\beta$), it is unknown if condition \eqref{eq:th1} is necessary for uniform stability when the
  class of $\Delta$'s is restricted to be memoryless. 
\end{rem}
In next subsection, we will show that when the set of $\Delta$'s is further confined, condition \eqref{eq:th1} is sufficient but not necessary.

\subsection{LTI uncertainty}
We consider here the case where $\Delta$'s are LTI systems.
\begin{lem}\label{le: LTI}
Given an LTI $\Delta \in \mathbf{\Delta_0}$, $ \Delta \in \mathbf{\Delta}_\infty$ if and only if $\Delta$ is a  nonnegative constant.
\end{lem}
\pf
``$\Leftarrow$''  When $\Delta$ is a nonnegative constant, it is clear that $\Delta$ is also monotone nondecreasing.  Then, it follows from Lemma
  \ref{le: static_1} that $\Delta\in \mathbf{\Delta}_\infty$.

  ``$\Rightarrow$'' Since $\Delta$ is a bounded causal LTI system on $\LT[0,\infty)$, it admits a transfer function representation $\hat{\Delta}(\cdot)$. 
By the Plancherel theorem, the inequality in \eqref{eq:58} can be expressed in the frequency domain as
\[
\begin{array}{rl}
 & \RE \left\lbrace \int_{-\infty}^{\infty}\hat{x}(j\omega)^* e^{j\omega\tau}\hat{\Delta}(j\omega)\hat{x}(j\omega)d\omega \right\rbrace\\
 \le & \RE\left\lbrace \int_{-\infty}^{\infty}  \hat{x}(j\omega)^* \hat{\Delta}(j\omega) \hat{x}(j\omega) d \omega\right\rbrace\\
  &  \quad\forall \tau\in\mathbb{R}, \forall x\in \Ltwo[0, \infty),
\end{array}
\]
which implies for all $\tau\in\mathbb{R}, \omega \in\mathbb{R}$
\begin{equation} \label{eq: condition of constant Delta}
\RE\left\lbrace \hat{\Delta}(j\omega)(1+e^{j\omega\tau})\right\rbrace\ge 0.
\end{equation} 
As the phase of $(1+e^{j\omega\tau})$ can take any value in $(-\frac{\pi}{2}, \frac{\pi}{2})$ at $\omega \neq 0$, $\hat{\Delta}(j\omega)$ would need
to be zero or have zero phase in order to preserve the nonnegativity in \eqref{eq: condition of constant Delta}.  In other words, $\Delta$ has to be a
nonnegative constant. $\hfill$\qed
 \endpf

Define 
\[
\mathbf{\Delta_{LTI}} := \left\lbrace \Delta \in \mathbf{\Delta}_\infty \; | \; \Delta  \text{ is LTI}   \right\rbrace.
\]
By Lemma~\ref{le: LTI}, $\mathbf{\Delta_{LTI}} = \{\Delta \in \Real \;|\; \Delta \geq 0\}$.
\begin{thm}\label{th: LTI}
  Consider Figure \ref{fig:1} with $G\in \mathcal{RH}_{\infty}$ and $\Delta \in \mathbf{\Delta_{LTI}}$, and suppose $[G,\Delta]$ is well-posed for
  all $\Delta \in \mathbf{\Delta_{LTI}}$.  Then, $[G,\Delta]$ is uniformly stable over $ \mathbf{\Delta_{LTI}}$ if \eqref{eq:th2} holds. Moreover, the converse is not true.
\end{thm}
\pf
  The first part follows from Corollary \ref{cor: static_2} since $\mathbf{\Delta_{LTI}}\subset \mathbf{\bar{\Delta}_{static}}$ by Lemma~\ref{le:
    LTI}. To prove that the converse is not true, consider the following O'Shea counterexample adapted from \cite[(42)]{wang2017phase}. Let
  $H(s) := -\frac{s^2}{(s^2+2\xi s+1)^2}$ and $G := H - \varepsilon \in \mathcal{RH}_{\infty}$, where $\xi\in (0,0.25]$ and $\varepsilon > 0$ is
  sufficiently small. Since the Nyquist plot of $G$ does not intersect with the nonnegative real line, it follows from the Nyquist stability theorem
  that $[G,\Delta]$ is uniformly stable over all nonnegative constant $\Delta$. By applying the results about the phase limitation on the Zames-Falb
  multipliers in \cite[Section III.C]{wang2017phase}, it follows that there exists no $z\in \Lone$ such that $\|z\|_1 \le 1$ and \eqref{eq:th1} holds
  true. $\hfill$\qed
\endpf

\begin{figure*}
\begin{center}
\includegraphics[height=9cm]{{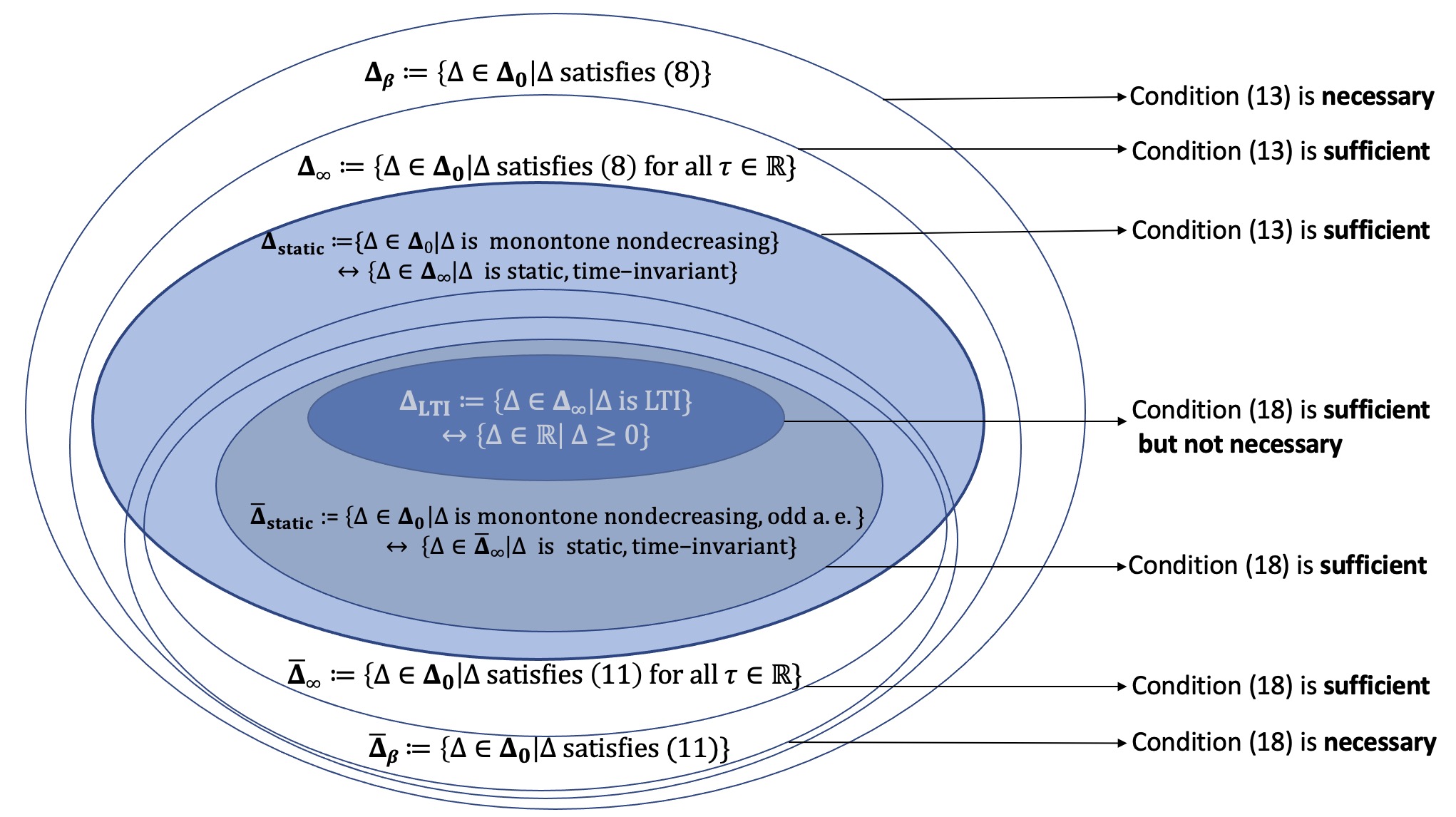}}    
\caption{Sufficiency \& necessity of conditions for robust uniform stability of $[G,\Delta]$ over sets of $\Delta$'s. }  
\label{fig: main results}                                 
\end{center}                                 
\end{figure*}

The results presented in this section are concluded in Figure \ref{fig: main results}.  As shown in the diagram, there is a subtle trade-off between the size of the class of $\Delta$'s and the strictness of
the condition on $G$ under which the robust feedback stability is guaranteed. 

\section{Extensions to Slope-Restricted  Uncertainty}\label{sec:4}
In this section, we continue investigating the uniform stability of the feedback system in Figure \ref{fig:1} with a smaller set of $\Delta$'s by
taking into account an additional `slope' restriction. The restriction is parameterized by the pair $(a,b)$ with $0\le a <b \le
\infty$.

\subsection{Nonlinear dynamic uncertainty}
Let $y(t)=(\Delta x)(t)$ and consider the following inequality on the operator $\Delta$:
\begin{equation}\label{eq:58_3}
\begin{array}{rl}
 &  \displaystyle \int_{0}^{\infty} \left[x-b^{-1}y\right](t+\tau) \left[-ax+y\right](t) dt \\
 \le &  \displaystyle  \int_{0}^{\infty}\left[ x-b^{-1}y\right](t)\left[-ax+y\right](t)dt, \\
   & \forall \tau\in (-\beta,-1/\beta) \cup (1/\beta,\beta),\forall x\in \LT[0,\infty)
\end{array}
\end{equation}
for some constants $b>a\ge0$ and $\beta\in(1,\infty]$. Also let
\begin{equation}
\mathbf{\Delta^{[a,b]}_\beta} := \left\lbrace \Delta \in \mathbf{\Delta_0} \; |\;\Delta\text{ satisfies } \eqref{eq:58_3} \right\rbrace.
\end{equation}


 Before presenting the main theorem, we establish some supporting lemmas. 
\begin{lem}\label{le: slope_1}
Given a $\Delta\in \mathbf{\Delta_0}$,  $\Delta$ satisfies \eqref{eq:58_3}  if and only if 
\begin{equation}\label{eq:62_2}
\left\langle x-b^{-1}y, (1-\mathbf{Z})(-ax+y)\right\rangle \ge 0, \forall x\in \LT[0,\infty)
\end{equation}
with  $y=\Delta x$ for all $z\in \Lone^{+}(\beta)$ such that $\|z\|_1 = 1$.
\end{lem}
\pf
This can be proved by rewriting  \eqref{eq:62_2} as
$
\left\langle x-b^{-1}y, \mathbf{Z}(-ax+y)\right\rangle \le \left\langle x-b^{-1}y, (-ax+y)\right\rangle
$ and  by using the same arguments in the proof of Lemma \ref{le:1}. 
 
 $\hfill$\qed
\endpf

Let $M(j\omega)$ denote $1-Z(j\omega)$. Lemma \ref{le: slope_1} shows that given a $\Delta \in \mathbf{\Delta_0}$, a necessary and sufficient condition for
$\Delta\in \mathbf{\Delta^{[a,b]}_\beta}$ is that $\Delta$ satisfies the class of IQCs  defined by
\begin{equation}\label{eq:IQC_slope}
 \begin{array}{l}
  \quad  \Pi^{[a,b]}(j\omega)\\
 = \left[ \begin{array}{cc}
 -a\left(  M(j\omega)+ M(j\omega)^{*}\right) & ab^{-1} M(j\omega)+ M(j\omega)^{*}\\
 ab^{-1} M(j\omega)^{*}+M(j\omega) & -b^{-1}\left(  M(j\omega)+ M(j\omega)^{*}\right)
 \end{array}\right]
\end{array}
\end{equation}  
for all $z\in \Lone^{+}(\beta)$ such that $\|z\|_1\le 1$. 

\begin{lem}\label{le:homotopy}
If $\Delta \in \mathbf{\Delta^{[a,b]}_\beta}$, then for all $\theta \in[0,1]$,
$
\theta \Delta +(1-\theta)a \in  \mathbf{\Delta^{[a,b]}_\beta}.
$
\end{lem}
\pf
  Given any $\Delta \in \mathbf{\Delta^{[a,b]}_\beta}$, \eqref{eq:58_3} holds by definition, and note that
  $\theta \Delta +(1-\theta)a \in \mathbf{\Delta_0}$ for all $\theta \in[0,1]$. It remains to verify that $\theta \Delta +(1-\theta)a$ also satisfies
  \eqref{eq:58_3} for any $\theta\in [0,1]$. Thus, it suffices to show that  \eqref{eq:58_3} with $y$ replaced by  $ \theta y +(1-\theta)ax$ holds true  for all $\theta \in[0,1]$.
 By  replacing $y$ with  $ \theta y +(1-\theta)ax$ in \eqref{eq:58_3}, it can be
  obtained that the inequality is equivalent to
\begin{equation}\label{eq:proof homotopy}
\begin{array}{rl}
 &\int_0^{\infty}  \left[ (x-b^{-1}y)+(1-\theta)b^{-1}(-ax+y)\right] (t+\tau)
 \\   & \quad\quad\cdot \left[ \theta(-ax+y) \right](t) dt\\
 \le & \int_0^{\infty}  \left[ (x-b^{-1}y)+(1-\theta)b^{-1}(-ax+y)\right] (t) \\
  & \quad\quad \cdot\left[ \theta(-ax+y) \right](t) dt\\
  & \qquad \forall \tau\in (-\beta,-1/\beta) \cup (1/\beta,\beta),\forall x\in \LT[0,\infty).
\end{array}
\end{equation}
To show that \eqref{eq:proof homotopy} holds, observe that
\begin{equation} \label{eq:lan_parseval}
\begin{array}{rl}
 & \int_{0}^{\infty}\left[-ax+y\right](t+\tau) \cdot \left[-ax+y\right](t)dt \\
 \le & \int_{0}^{\infty}\left[-ax(t)+y(t)\right]^{2}dt\\
  & \quad \forall \tau\in (-\beta,-1/\beta) \cup (1/\beta,\beta),\forall x\in \LT[0,\infty),
\end{array}
 \end{equation}
which follows from the fact that
\[
\begin{array}{rl}
 & \int_{-\infty}^{\infty}e^{j\omega \tau} \left[-a\hat{x}(j\omega)+\hat{y}(j\omega)\right]^* \left[-a\hat{x}(j\omega)+\hat{y}(j\omega)\right]d\omega \\
 \le & \int_{-\infty}^{\infty}\left[-a\hat{x}(j\omega)+\hat{y}(j\omega)\right]^*\left[-a\hat{x}(j\omega)+\hat{y}(j\omega)\right] d\omega\\
  & \quad \forall \tau\in (-\beta,-1/\beta) \cup (1/\beta,\beta),\forall x\in \LT[0,\infty)
\end{array}
 \]
 and the Parseval's theorem. Then, multiply \eqref{eq:lan_parseval} and \eqref{eq:58_3} by $\theta(1-\theta)b^{-1}$ and $\theta$ respectively, and
 \eqref{eq:proof homotopy} follows by adding up the two obtained inequalities. $\hfill$\qed
\endpf
Next, consider the  inequality 
\begin{equation}\label{eq:58_4}
 \begin{array}{rl}
 & \displaystyle \left|\int_{0}^{\infty} \left[x-b^{-1}y\right](t+\tau)\left[-ax+y\right](t) dt \right|\\
 \le & \displaystyle \int_{0}^{\infty}\left[ x-b^{-1}y\right](t)\left[-ax+y\right](t)dt, \\
   & \quad \forall \tau\in (-\beta,-1/\beta) \cup (1/\beta,\beta),\forall x\in \LT[0,\infty)
\end{array}
\end{equation}
and define
\begin{equation}
\mathbf{\bar{\Delta}^{[a,b]}_\beta} := \left\lbrace \Delta  \in \mathbf{\Delta_0} \; |\; \Delta \text{  satisfies } \eqref{eq:58_4} \right\rbrace.
\end{equation}
\begin{lem} \label{le: slope_odd}
Given a $\Delta \in \mathbf{\Delta_0}$,  $\Delta$ satisfies \eqref{eq:58_4} if and only if \eqref{eq:62_2} holds
with   $y=\Delta x$ for all  $z\in \Lone(\beta)$ such that $\|z\|_1\le 1$. If $\Delta \in \mathbf{\bar{\Delta}^{[a,b]}_\beta}$, then for all $\theta \in[0,1]$,
$
\theta \Delta +(1-\theta)a \in  \mathbf{\bar{\Delta}^{[a,b]}_\beta}.
$
\end{lem}

The proof to the lemma above is omitted since it follows the same lines of reasoning in  Lemma \ref{le: 1_2} and Lemma \ref{le:homotopy}.

A variant of the IQC stability theorem tailored to our purpose is stated next.

\begin{prop}\label{pro:1}
Given  $\Pi \in \mathcal{L}_{\infty}, G\in \mathcal{RH}_{\infty}$ and bounded causal $\Delta$, assume that 
\begin{enumerate} \renewcommand{\theenumi}{\textup{(\roman{enumi})}}\renewcommand{\labelenumi}{\theenumi}
\item for $\theta \in [0,1]$ the interconnection $[G,\theta \Delta +(1-\theta)a]$ is well-posed,
\item the interconnection $[G,a]$ is stable,
\item for $\theta \in [0,1]$, $\theta \Delta +(1-\theta)a \in \text{IQC}(\Pi)$,
\item there exists $\epsilon>0$ such that 
\[
\left[\begin{matrix}
G(j\omega)\\ 1
\end{matrix}\right]^*
\Pi(j\omega)
\left[\begin{matrix}
G(j \omega)\\ 1
\end{matrix}\right]\le -\epsilon, \quad\forall \omega  \in\Real.
\]
\end{enumerate}
Then the system  $[G,\Delta]$ is stable.
\end{prop}
\pf The result can be established using the same line of reasoning of \cite[Theorem 1]{megretski1997system},
$\hfill$\qed
\endpf

\begin{thm}\label{thm:slope}
Consider Figure \ref{fig:1} with $G\in \mathcal{RH}_{\infty}$. Suppose that $[G,a]$ is stable and $[G,\Delta]$ is well-posed for all $\Delta \in \mathbf{\Delta^{[a,b]}_\infty}$
  (respectively, $\Delta \in \mathbf{\bar{\Delta}^{[a,b]}_\infty}$).  Then, $[G,\Delta]$ is uniformly stable over $\mathbf{\Delta^{[a,b]}_\infty}$ (respectively,
  $\mathbf{\bar{\Delta}^{[a,b]}_\infty}$) if  
\begin{equation} \label{eq:slope IQC theorem}
\begin{array}{l}
 \exists z\in \Lone^{+}(-\infty,\infty) (\text{respectively, }\Lone (-\infty,\infty)),\epsilon>0  \\
 \text{ s.t. }  \quad \|z\|_1\le 1 \text{ and  }  \forall \omega\in \Real,\\ 
     \RE  \left\lbrace (1-Z(j\omega))(G(j\omega)-b^{-1}) (aG(j\omega)^{*}-1)\right\rbrace \ge  \epsilon.
\end{array}
\end{equation}
Moreover, $[G,\Delta]$ is uniformly stable over $\mathbf{\Delta^{[a,b]}_\beta}$ (respectively,
  $\mathbf{\bar{\Delta}^{[a,b]}_\beta}$) for any $\beta \in (1, \infty)$ only if \eqref{eq:slope IQC theorem} holds.
\end{thm}
\pf
First consider the case where $\Delta \in \mathbf{\Delta^{[a,b]}_\beta}$.

``$\Leftarrow$''  By letting $\beta=\infty$,  we have from Lemma \ref{le: slope_1} and  Lemma \ref{le:homotopy} that for each $\Delta\in \mathbf{\Delta^{[a,b]}_\infty}$,
$\theta \Delta +(1-\theta)a \in \mathbf{\Delta^{[a,b]}_\infty} $ for $\theta \in [0,1]$, which implies (iii) in Proposition \ref{pro:1} with $\Pi$ defined by
\eqref{eq:IQC_slope} with  all $z\in \Lone^{+}(-\infty,\infty)$ such that $\|z\|_1\le 1$ holds for all $\Delta\in \mathbf{\Delta^{[a,b]}_\infty}$. Moreover, it also implies that (i) in Proposition \ref{pro:1} holds for all
$\Delta\in \mathbf{\Delta^{[a,b]}_\infty}$.  Now observe that condition \eqref{eq:slope IQC theorem} can be equivalently transformed into
\[
\left[\begin{matrix}
G (j\omega)\\ 1
\end{matrix}\right]^*
 \Pi^{[a,b]}(j\omega)
\left[\begin{matrix}
G (j\omega)\\ 1
\end{matrix}\right]\le -2\epsilon , \forall \omega \in \Real.
\]
By exploiting Proposition \ref{pro:1}, it can be concluded that $[G,\Delta]$ with $\Delta \in\mathbf{\Delta^{[a,b]}_\infty}$ is robustly stable. The
uniform stability of $[G,\Delta]$ can be further established using the same arguments  in the proof of Theorem \ref{th:1}.

``$\Rightarrow$''  The necessity can be proved following the same line of the necessity proof in Theorem \ref{th:1}.

The case of  $\Delta \in \mathbf{\bar{\Delta}^{[a,b]}_\beta}$ can be proved analogously based on Lemma \ref{le: slope_odd}.  $\hfill$\qed
\endpf
\begin{rem}
 It can be easily verified that when $a=0$ and $b\rightarrow\infty$, Theorem  \ref{thm:slope} specializes to Theorems \ref{th:1} and \ref{th:1_2}.  We note that it is challenging to show the sufficiency of Theorem  \ref{thm:slope}  by applying    loop-transformation techniques to the results in Section \ref{sec:3}.  The difficulties lie in the unboundedness of  $\tilde{\Delta} :=(\Delta-a)(b-\Delta)^{-1}$ or showing $\mathbf{\Delta^{[a,b]}_\beta}\subset \mathbf{\Delta^{[a,b+\delta]}_\beta}$ where $\delta$ is some arbitrarily small positive number. 
\end{rem}

\subsection{Static uncertainty}
Next, we consider the class of static (i.e. memoryless) $\Delta$'s in the set $\mathbf{\Delta^{[a,b]}_\beta}$ (respectively,
$\mathbf{\bar{\Delta}^{[a,b]}_\beta}$). The following lemma explains the links between these sets with certain types of slope-restricted static nonlinearity
that are widely considered. Note that the necessity directions in the next lemma cannot be obtained by applying loop
  transformations~\cite{zames1967stability} to the $\Delta$ in Lemmas~\ref{le: static_1} and~\ref{le: static_2}, since doing so would result in
  unbounded operators. As such, a different construction to those in Lemmas~\ref{le: static_1} and~\ref{le: static_2} is needed.
\begin{lem} \label{le: slope_static}
Given a static and time-invariant  $\Delta \in \mathbf{\Delta_0}$, then
\begin{enumerate} \renewcommand{\theenumi}{\textup{(\roman{enumi})}}\renewcommand{\labelenumi}{\theenumi}
\item $\Delta\in \mathbf{\Delta^{[a,b]}_\infty}$ if and only if  
\begin{equation}\label{eq: slope-restriction}
a\le \frac{\Delta(x_1)-\Delta(x_2)}{x_1-x_2} \le b, \forall x_1,x_2\in\mathbb{R},x_1 \neq x_2.
\end{equation}
\item  $\Delta\in \mathbf{\bar{\Delta}^{[a,b]}_\infty}$  if and only if  $\Delta$ is odd and satisfies \eqref{eq: slope-restriction}.
\end{enumerate}
\end{lem}
\pf
Throughout the proof, let $\bar{x} := x-b^{-1}\Delta (x)$ and $\bar{y} := -ax+\Delta(x)$. 

(i)  Note that  $\Delta\in \mathbf{\Delta^{[a,b]}_\infty}$ can be rewritten as
\begin{equation}\label{eq:58_5}
\begin{split}
\int_{0}^{\infty} \bar{x}(t+\tau)\bar{y}(t)dt \le \int_{0}^{\infty} \bar{x}(t)\bar{y}(t)dt \\
\forall \tau\in \Real, x\in \LT[0,\infty).\qquad
\end{split}
\end{equation} 
``$\Rightarrow$''  Given any $x_1,x_2 \in \Real$ such that $x_1<x_2$. If $\bar{x}_1=\bar{x}_2$, it follows that $\Delta(x_1)-\Delta(x_2)=b(x_1-x_2)$,
which satisfies \eqref{eq: slope-restriction}.  If $\bar{x}_1\neq\bar{x}_2$, define $x(t) \in \mathcal{L}_{2} [0,\infty)$ as in \eqref{eq:nec_ex}
for some $L\in \mathbb{Z}^+$.  Then it follows that
\[
\begin{array}{rl}
 & \bar{x}(t+1)\bar{y}(t) - \bar{x}(t)\bar{y}(t) \\
 = & \left\lbrace
 \begin{array}{ll} \bar{x}_2\bar{y}_1-\bar{x}_1\bar{y}_1 & \text{if} \;  t\in [0,1)\cup\cdots \cup [2L,2L+1)\\
\bar{x}_1\bar{y}_2-\bar{x}_2\bar{y}_2& \text{if} \;t\in [1,2) \cup \cdots \cup [2L-1,2L)\\
 -\bar{x}_2\bar{y}_2 &   \text{if} \; t\in [2L+1,2L+2)\\
 0 & \text{otherwise}.
\end{array} \right.
\end{array}
\]
To satisfy \eqref{eq:58_5} when $\tau=1$ and $L$ is  sufficiently large,  it is thus required that 
\begin{equation}\label{eq: hatxy}
\bar{x}_2\bar{y}_1-\bar{x}_1\bar{y}_1+\bar{x}_1\bar{y}_2-\bar{x}_2\bar{y}_2\le 0.
\end{equation}
If $\bar{x}_1>\bar{x}_2$, then $ \frac{\Delta(x_1)-\Delta(x_2)}{x_1-x_2} > b, $ and \eqref{eq: hatxy} implies that $\bar{y}_1\ge\bar{y}_2$, which
gives $ \frac{\Delta(x_1)-\Delta(x_2)}{x_1-x_2} \le a.  $ This gives rise to a contradiction since $b>a$. Hence, it must hold that
$\bar{x}_1 < \bar{x}_2$. In this case, it follows from \eqref{eq: hatxy} that $\bar{y}_1\le\bar{y}_2$. Together, they imply
$ a \le \frac{\Delta(x_1)-\Delta(x_2)}{x_1-x_2} < b.  $ Consequently, \eqref{eq: slope-restriction} follows.

``$\Leftarrow$''  Suppose $ a \le \frac{\Delta(x_1)-\Delta(x_2)}{x_1-x_2} < b $. Then it is proved in \cite[Section 7]{zames1967stability} that the
function mapping $\bar{x}$ to $\bar{y}$ is static, bounded and monotone nondecreasing. Therefore it follows from Lemma \ref{le: static_1} that
\eqref{eq:58_5} holds true. Note that the set of $\Delta$'s satisfying \eqref{eq: slope-restriction} is the closure of the set of $\Delta$'s satisfying
$a \le \frac{\Delta(x_1)-\Delta(x_2)}{x_1-x_2} < b$. It is then straightforward to see that \eqref{eq:58_3} holds for all $\Delta$ satisfying \eqref{eq:
  slope-restriction} with $\beta=\infty$ via a limiting argument.

(ii) ``$\Rightarrow$'' Given a static $\Delta\in \mathbf{\bar{\Delta}^{[a,b]}_\infty}$, since $\mathbf{\bar{\Delta}^{[a,b]}_\infty}\subset\mathbf{\Delta^{[a,b]}_\infty}$, it
follows from (i) that $\Delta$ satisfies \eqref{eq: slope-restriction}, whereby $\Delta$ is continuous. It remains to show that $\Delta$ is odd.
First, note from \eqref{eq: slope-restriction} that both $x-b^{-1}\Delta(x)$ and $-ax+\Delta(x)$ are monotone nondecreasing.

If there exist $x_1,x_2$ such that $0<\bar{x}_{1}<\bar{x}_2$, then one has $0<x_1<x_2$ and $0<\bar{y}_1\le \bar{y}_2$. In this case, let
$\bar{x}_{-1}=-x_1-b^{-1}\Delta(-x_1), \bar{x}_{-2}=-x_2-b^{-1}\Delta(-x_2)$ and $\bar{y}_{-1}=ax_1+\Delta(-x_1), \bar{y}_{-2}=ax_2+\Delta(-x_2)$. By
defining $x(t)$ as in \eqref{eq: construct x_1} and \eqref{eq: construct x_2} respectively for a sufficiently large $L\in \mathbb{Z}^+$ and
investigating the term $\bar{x}(t+1)\bar{y}(t) + \bar{x}(t)\bar{y}(t)$, it can be shown using the same argument of the necessity proof of Lemma
\ref{le: static_2} that $\bar{y}_1=-\bar{y}_{-1}$. As a result, $-ax_1+\Delta(x_1)=-ax_1-\Delta(-x_1)$, which gives $\Delta(x_1)=-\Delta(-x_1)$. In
other words, $\Delta$ is odd.

If there exist no $x_1,x_2$ such that $0<\bar{x}_1<\bar{x}_2$, then it can be implied that $\Delta(x)=bx$, and hence $\Delta$ is odd.

``$\Leftarrow$''  This can be shown based on the sufficiency  proof for (i) and Lemma \ref{le: static_2}. $\hfill$\qed
\endpf

Define 
\begin{equation} \label{eq: static_slope_set of Delta}
\mathbf{\Delta_{static}^{[a,b]}} := \left\lbrace \Delta  \in \mathbf{\Delta_0} \; | \;  \Delta \text{ is static and satisfies }  \eqref{eq: slope-restriction} \right\rbrace
\end{equation}
and
\begin{equation} \label{eq: static_slope_set of odd Delta} \mathbf{\bar{\Delta}_{static}^{[a,b]}} := \left\lbrace \Delta \in
    \mathbf{\Delta_{static}^{[a,b]}} \; | \;  \Delta \text{ is odd }  \right\rbrace.
\end{equation}
By Lemma \ref{le: slope_static},
\[\mathbf{\Delta_{static}^{[a,b]}}= \left\lbrace \Delta\in \mathbf{\Delta^{[a,b]}_\infty} \; |\; \Delta \text{ is static} \allowbreak \text{ and  time-invariant} \right\rbrace \] and
\[\mathbf{\bar{\Delta}_{static}^{[a,b]}}= \left\lbrace \Delta\in \mathbf{\bar{\Delta}^{[a,b]}_\infty} \; |\; \Delta \text{ is static and time-invariant} \right\rbrace .\] 
Hence,
$\mathbf{\Delta_{static}^{[a,b]}} \subset \mathbf{\Delta^{[a,b]}_\infty}$ and
$\mathbf{\bar{\Delta}_{static}^{[a,b]}}\subset\mathbf{\bar{\Delta}^{[a,b]}_\infty}$. This together with Theorem \ref{thm:slope} enables us to provide the
following result.

\begin{cor}\label{cor: static_slope}
  Consider Figure \ref{fig:1} with $G\in \mathcal{RH}_{\infty}$ and $\Delta \in \mathbf{\Delta_{static}^{[a,b]}}$ (respectively,
  $\mathbf{\bar{\Delta}_{static}^{[a,b]}}$). Suppose that $[G,a]$ is stable and $[G,\Delta]$ is well-posed for all
  $\Delta \in \mathbf{\Delta_{static}^{[a,b]}}$ (respectively, $\Delta \in \mathbf{\bar{\Delta}_{static}^{[a,b]}}$).  Then, $[G,\Delta]$ is uniformly
  stable over $\mathbf{\Delta_{static}^{[a,b]}}$ (respectively, $ \mathbf{\bar{\Delta}_{static}^{[a,b]}}$) if  \eqref{eq:slope IQC theorem} holds.
\end{cor}

In next subsection, we will show that when the set of $\Delta$'s is further confined, condition \eqref{eq:slope IQC theorem} is sufficient but
not necessary.
\subsection{LTI uncertainty}

Lastly we consider the case where $\Delta$'s are LTI systems.  
\begin{lem} \label{le: slope_LTI}
Given an LTI $\Delta \in \mathbf{\Delta_0}$,  $\Delta\in \mathbf{\Delta}^{[a,b]}_\infty$ if and only if  $\Delta$ is  a constant in $[a,b]$.
\end{lem}
\pf The sufficiency follows from \ref{le: slope_static} and the fact that a constant $\Delta$ in $[a,b]$ satisfies \eqref{eq: slope-restriction}. The
necessity can be shown following the same line of argument in the proof of Lemma \ref{le: LTI} with $\hat{\Delta}(j\omega)$ in \eqref{eq: condition of
  constant Delta} replaced by $(1-b^{-1}\hat{\Delta}(j\omega))(-a+\hat{\Delta}(j\omega)^{*})$. In particular, it can be verified that
$(1-b^{-1}\hat{\Delta}(j\omega))(-a+\hat{\Delta}(j\omega)^{*})$ is nonnegative real-valued, from which it follows that $\hat{\Delta}(j\omega)$ is
$[a,b]$-valued, which in turn implies that $\hat{\Delta}$ is a constant in $[a,b]$. $\hfill$\qed
\endpf
Define 
\[
\mathbf{\Delta_{LTI}^{[a,b]}} := \left\lbrace \Delta \in \mathbf{\Delta^{[a,b]}_\infty} \; | \;   \Delta  \text{ is  LTI} \right\rbrace.
\]
By Lemma \ref{le: slope_LTI}, $\mathbf{\Delta_{LTI}^{[a,b]}} =\left\lbrace \Delta \in \Real  \; | \;  a \le \Delta\le b  \right\rbrace.$
\begin{cor}
  Consider Figure \ref{fig:1} with $G\in \mathcal{RH}_{\infty}$ and $\Delta \in \mathbf{\Delta_{LTI}^{[a,b]}}$. Suppose that $[G,a]$ is stable and
  $[G,\Delta]$ is well-posed for all $\Delta \in \mathbf{\Delta_{LTI}^{[a,b]}}$. Then, $[G,\Delta]$ is uniformly stable over
  $\mathbf{\Delta_{LTI}^{[a,b]}}$ if \eqref{eq:slope IQC
    theorem} holds. Moreover, the converse is not true.
\end{cor}
\pf The sufficiency can be proved in a straightforward manner with Corollary \ref{cor: static_slope} since
$\mathbf{\Delta_{LTI}^{[a,b]}} \subset \mathbf{\bar{\Delta}_{static}^{[a,b]}}$.  To prove the converse is not true, first observe that condition
\eqref{eq:slope IQC theorem} can be rewritten as
\begin{equation} \label{eq: IQC_2}
\begin{array}{rl}
 \RE & \left\lbrace(1-Z(j\omega))(G(j\omega)-b^{-1}) \right. \\
       & \left. (aG(j\omega)-1)^{-1}\right\rbrace \ge \hat{\epsilon} \quad \forall \omega\in \Real
\end{array}
\end{equation}
for some $\hat{\epsilon}>0$, in which the stability of $[G, a]$ ensures that $(aG - 1)^{-1} \in \mathcal{RH}_{\infty}$.

If $a=0$, let $-G(s)=\frac{s^2}{(s^2+2\xi s+1)^2}+\varepsilon -b^{-1}$, where $\xi \in (0,0.25]$ and $\varepsilon>0$ is sufficiently small. Then we
have $G\in \mathcal{RH}_{\infty}$ and $(G(s)-b^{-1})(aG(s)-1)^{-1}=\frac{s^2}{(s^2+2\xi s+1)^2}+\varepsilon$. Since the Nyquist plot of
$\frac{s^2}{(s^2+2\xi s+1)^2}+\varepsilon$ does not intersect with nonpositive real line, it can be implied from the above equation that the Nyquist plot
of $G$ does not intersect the interval $[b^{-1},a^{-1}]$. This, together with the fact that $[G, a]$ is stable, indicates that $[G,\Delta]$
is uniformly stable over $\mathbf{\Delta_{LTI}^{[a,b]}}$. Following the same line of reasoning in the proof of Theorem \ref{th: LTI}, it can be
shown that there is no $z\in \Lone$ such that $\|z\|_1\le 1$ and \eqref{eq: IQC_2} holds true.

If $a>0$, let $G(s)=\frac{(b^{-1}-\varepsilon)(s^2+2\xi s+1)^2-a^{-1}\xi^{2}s^2}{(1-a\varepsilon)(s^2+2\xi s+1)^2-\xi ^{2}s^2}$, where
$\xi \in (0,0.25]$ and $\varepsilon>0$ is sufficiently small. One can verify that $G\in \mathcal{RH}_{\infty}$ by, say, the Routh-Hurwitz stability
criterion and that $(G(s)-b^{-1})(aG(s)-1)^{-1}=\frac{a^{-1}\xi ^{2}s^2}{(s^2+2\xi s+1)^2}+\varepsilon$. Similarly, since the Nyquist plot of
$\frac{a^{-1}\xi ^{2}s^2}{(s^2+2\xi s+1)^2}+\varepsilon$ does not intersect with nonpositive real line, it can be implied from the above equation that
the Nyquist plot of $G$ does not intersect the interval $[b^{-1},a^{-1}]$. This and the fact that $[G, a]$ is stable imply that $[G,\Delta]$ is
uniformly stable over $\mathbf{\Delta_{LTI}^{[a,b]}}$. We note that $\frac{a^{-1}\xi ^{2}s^2}{(s^2+2\xi s+1)^2}$ differs from
\cite[(42)]{wang2017phase} only by a positive factor and hence they have the same phase response. By applying the results on phase limitation in
\cite[Section III.C]{wang2017phase}, we can again conclude that there is no $z\in \Lone$ such that $\|z\|_1\le 1$ and \eqref{eq: IQC_2} holds
true. $\hfill$\qed
\endpf

\section{Conclusion}\label{sec:5}
In the systems and control literature, the Zames-Falb multipliers are widely used as a classical tool to establish the input-output stability of a
feedback interconnection of an LTI system and a static monotone nonlinearity. Not much attention has been paid to investigating the conservatism of
using the Zames-Falb multipliers. 
This paper identifies a class of uncertain systems
over which the robust feedback stability implies the existence of an appropriate Zames-Falb multiplier based on the generalised S-procedure lossless theorem. Meanwhile, it is shown that the existence of such a Zames-Falb multiplier is sufficient for the robust feedback stability over a smaller class of uncertain systems. When restricted to be
 static (a.k.a. memoryless), this class of systems coincides with the class of monotone nonlinearities  with possible slope-restrictions considered in
the classical paper~\cite{zames1967stability}, and the classical result
of using the Zames-Falb multipliers to ensure feedback stability is recovered.  When the same class of systems is restricted to be LTI,  the existence of a Zames-Falb multiplier is shown to be sufficient but not necessary for the robust feedback
stability. Nevertheless it remains unknown whether the existence of a Zames-Falb multiplier is necessary for the uniform feedback stability over these classes. This gives rise to an interesting future research direction. 

\begin{ack}
  The authors gratefully acknowledge Peter Seiler and Andrey Kharitenko for many useful discussions.
\end{ack}
\bibliographystyle{plain} \bibliography{zf_reference}

\begin{thebibliography}{10}

\bibitem{Bre11}
H.~Brezis.
\newblock {\em Functional Analysis, Sobelev Spaces, and Partial Differential
  Equations}.
\newblock Springer, 2011.

\bibitem{carrasco2013equivalence}
J.~Carrasco, W.~P. Heath, and A.~Lanzon.
\newblock Equivalence between classes of multipliers for slope-restricted
  nonlinearities.
\newblock {\em Automatica}, 49(6):1732--1740, 2013.

\bibitem{carrasco2014lmi}
J.~Carrasco, M.~Maya-Gonzalez, A.~Lanzon, and W.~P. Heath.
\newblock {LMI} searches for anticausal and noncausal rational {Z}ames{--F}alb
  multipliers.
\newblock {\em Systems and Control Letters}, 70:17--22, 2014.

\bibitem{carrasco2016zames}
J.~Carrasco, M.~C. Turner, and W.~P. Heath.
\newblock Zames-{F}alb multipliers for absolute stability: {F}rom
  {O}{'}{S}hea{'}s contribution to convex searches.
\newblock {\em European Journal of Control}, 28:1--19, 2016.

\bibitem{curtain2012introduction}
R.~F. Curtain and H.~Zwart.
\newblock {\em An introduction to infinite-dimensional linear systems theory},
  volume~21.
\newblock Springer Science \& Business Media, 1995.

\bibitem{Doy84}
J.~C. Doyle.
\newblock {\em Lecture Notes in Advances in Multivariable Control}.
\newblock ONR/Honeywell Workshop, Minneapolis, 1984.

\bibitem{dullerud2013course}
G.~E. Dullerud and F.~Paganini.
\newblock {\em A course in robust control theory: a convex approach},
  volume~36.
\newblock Springer Science \& Business Media, 2013.

\bibitem{johnson2015schauder}
W.~B. Johnson and G.~Schechtman.
\newblock A {S}chauder basis for {$L_{1}(0,\infty)$} consisting of non-negative
  functions.
\newblock {\em Illinois Journal of Mathematics}, 59(2):337--344, 2015.

\bibitem{Jonsson01lecturenotes}
U.~J{\" o}nsson.
\newblock Lecture notes on integral quadratic constraints.
\newblock Department of Mathematics, Royal Instutue of Technology (KTH),
  Stockholm, Sweden, 2001.

\bibitem{jonsson1996stability}
U.~J\"{o}nsson and M.~C. Laiou.
\newblock Stability analysis of systems with nonlinearities.
\newblock In {\em Proc. 35th IEEE Conf. Decision Control}, volume~2, pages
  2145--2150, 1996.

\bibitem{khalil2002nonlinear}
H.~K. Khalil.
\newblock {\em Nonlinear Systems}.
\newblock Prentice Hall, 3rd edition, 2002.

\bibitem{Kho22}
S.~Z. Khong.
\newblock On integral quadratic constraints.
\newblock {\em IEEE Trans. Autom. Contr.}, 2022.
\newblock In press.

\bibitem{KhoKao21}
S.~Z. Khong and C.-Y. Kao.
\newblock Converse theorems for integral quadratic constraints.
\newblock {\em IEEE Trans. Autom. Contr.}, 2021.
\newblock In press.

\bibitem{KhoSch18}
S.~Z. Khong and A.~van~der Schaft.
\newblock On the converse of the passivity and small-gain theorems for
  input-output maps.
\newblock {\em Automatica}, 97:58--63, 2018.

\bibitem{KulSaf02}
V.~V. Kulkarni and M.~G. Safonov.
\newblock All multipliers for repeated monotone nonlinearities.
\newblock {\em IEEE Trans. Autom. Contr.}, 47(7):1209--1212, 2002.

\bibitem{megretski1995combining}
A.~Megretski.
\newblock Combining $l_1$ and $l_2$ methods in the robust stability and
  performance analysis of nonlinear systems.
\newblock In {\em Proc. 34th IEEE Conf. Decision Control}, volume~3, pages
  3176--3181, 1995.

\bibitem{megretski1997system}
A.~Megretski and A.~Rantzer.
\newblock System analysis via integral quadratic constraints.
\newblock {\em IEEE Trans. Autom. Contr.}, 42(6):819--830, 1997.

\bibitem{MegTre93}
A.~Megretski and S.~Treil.
\newblock Power distribution inequalities in optimization and robustness of
  uncertain systems.
\newblock {\em J. Math. Syst., Estimat. Control}, 3(3):301--319, 1993.

\bibitem{o1966combined}
R.~O{'}shea.
\newblock A combined frequency-time domain stability criterion for autonomous
  continuous systems.
\newblock {\em IEEE Trans. Autom. Contr.}, 11(3):477--484, 1966.

\bibitem{o1967improved}
R.~O{'}shea.
\newblock An improved frequency time domain stability criterion for autonomous
  continuous systems.
\newblock {\em IEEE Trans. Autom. Contr.}, 12(6):725--731, 1967.

\bibitem{Pop61}
V.~M. Popov.
\newblock Absolute stability of nonlinear systems of automatic control.
\newblock {\em Automation and Remote Control}, 22:857--875, 1961.

\bibitem{SeiCar21}
P.~Seiler and J.~Carrasco.
\newblock Construction of periodic counterexamples to the discrete-time
  {Kalman} conjecture.
\newblock {\em IEEE Control Systems Letters}, 5(4):1291--1296, 2021.

\bibitem{toland2020dual}
J.~Toland.
\newblock {\em The Dual of $L_\infty(X, \mathcal{L}, \lambda)$, Finitely
  Additive Measures and Weak Convergence: A Primer}.
\newblock Springer Nature, 2020.

\bibitem{turner2019analysis}
M.~C. Turner and R.~Drummond.
\newblock Analysis of systems with slope restricted nonlinearities using
  externally positive zames{--F}alb multipliers.
\newblock {\em IEEE Trans. Autom. Contr.}, 65(4):1660--1667, 2019.

\bibitem{turner2009existence}
M.~C. Turner, M.~Kerr, and I.~Postlethwaite.
\newblock On the existence of stable, causal multipliers for systems with slope
  restricted nonlinearities.
\newblock {\em IEEE Trans. Autom. Contr.}, 54(11):2697--2702, 2009.

\bibitem{wang2017phase}
S.~Wang, J.~Carrasco, and W.~P. Heath.
\newblock Phase limitations of {Z}ames-{F}alb multipliers.
\newblock {\em IEEE Trans. Autom. Contr.}, 63(4):947--959, 2017.

\bibitem{Yak82}
V.~A. Yakubovich.
\newblock On an abstract theory of absolute stability of nonlinear systems.
\newblock {\em Vestnik Leningrad University Math.}, 10:341--361, 1982.
\newblock Russian originally published in 1977.

\bibitem{zames1967stability}
G.~Zames and P.~L. Falb.
\newblock On the stability of systems with monotone and odd monotone
  nonlinearities.
\newblock {\em IEEE Trans. Autom. Contr.}, 12(2):221--223, 1967.

\bibitem{ZCH20}
J.~Zhang, J.~Carrasco, and W.~Heath.
\newblock Duality bounds for discrete-time {Zames--Falb} multipliers.
\newblock Under consideration. URL: arxiv.org/abs/2009.11975.

\bibitem{ZDG96}
K.~Zhou, J.~C. Doyle, and K.~Glover.
\newblock {\em Robust and Optimal Control}.
\newblock Prentice-Hall, Upper Saddle River, NJ, 1996.

\end{thebibliography}

\end{document}